\documentclass[11pt]{article}
\usepackage[a4paper, margin=1in]{geometry}
\usepackage{titlesec}
\usepackage{graphicx}
\usepackage{amsmath}
\usepackage{amssymb}
\usepackage{booktabs}
\usepackage{hyperref}
\usepackage{subfig}
\usepackage[utf8]{inputenc}
\usepackage[T1]{fontenc}
\usepackage{textcomp}
\usepackage{float}
\usepackage{caption}
\usepackage{authblk}
\usepackage{tikz}
\usepackage{quantikz}
\usepackage{adjustbox}
\usepackage{siunitx}  
\usepackage{xcolor}   

\titleformat{\section}{\large\bfseries}{\thesection.}{1em}{}

\title{\textbf{Quantum Similarity-Driven QUBO Framework for Multi-Period Supply Chain Allocation using Time-Multiplexed Coherent Ising Machines and Simulated Quantum Annealing}}

\author[1]{Rushikesh Ubale}
\author[2]{Yasar Mulani}
\author[3,4]{Abhay Suresh}
\author[5]{Gregory Byrd}
\author[5]{Sangram Deshpande}
\author[6]{B. R. Nikilesh}
\author[7]{Sanya Nanda}

\affil[1]{Research Software Engineer, Qkrishi Quantum, India \\ \texttt{rushikesh.ubale@qkrishi.com}}
\affil[2]{Junior Research Intern, Qkrishi Quantum, India \\ \texttt{mulaniyhofficial@gmail.com}}
\affil[3]{Junior Research Intern, Qkrishi Quantum, India \\ \texttt{abhaysuresh0810@gmail.com}}
\affil[4]{QIC, Indian Institute of Technology (IIT) Jodhpur, India}
\affil[5]{Department of Electrical and Computer Engineering, North Carolina State University, USA \\ 
\texttt{gbyrd@ncsu.edu, ssdesh24@ncsu.edu}}
\affil[6]{Quantum Software Developer, Quanfluence Private Limited, India \\ \texttt{brnikilesh7@gmail.com}}
\affil[7]{IBM Qiskit Advocate, India \\ \texttt{sanya.nanda@gmail.com}}

\date{}
\renewenvironment{abstract}{%
    \begin{center}%
    \normalfont\large\bfseries Abstract\par
    \end{center}%
    \vspace{1em}%
    \normalfont\normalsize 
}{}

\newcommand{\keywords}[1]{%
  \vspace{0.5em}
  \noindent\textbf{\small Keywords:} \textit{\small #1}
}

\begin{document}

\maketitle

\begin{abstract}
Multi-period stock-keeping unit (SKU) allocation in supply chains is a combinatorial optimization problem that is both NP-hard and operationally critical, requiring joint consideration of profitability, capacity feasibility, and assortment diversity. Quadratic unconstrained binary optimization (QUBO) provides a principled framework for such problems, yet most prior demonstrations either rely on simplified benchmarks or restrict quantum processing to preprocessing roles without embedding full operational constraints.

This work introduces a \emph{hybrid QUBO framework} that combines three key advances: (i) a quantum-derived similarity kernel, obtained from a variational RX embedding, to discourage redundant SKU selections and promote diversity; (ii) exact per-period capacity enforcement via slack-bit encoding, preserving feasibility within a pure quadratic formulation; and (iii) execution on a time-multiplexed Coherent Ising Machine (CIM) with benchmarking against simulated quantum annealing (SQA) and classical optimization algorithms. The full QUBO comprised over one million terms and approximately 4,100 variables, capturing profit, risk, similarity, and inventory constraints in a unified formulation.

On a feature-engineered dataset of 500 SKUs over eight planning periods, Quanfluence’s CIM achieved a best-found energy of $-2.95\times 10^{16}$ and generated operationally robust solutions: 288 distinct SKUs ($\approx$60\%  of catalog), 226,813 allocated units, and a net profit of \$12.75M, with zero capacity violations. Importantly, the solver consistently enforced top-profit SKU presence while maintaining low redundancy. 

By integrating quantum-inspired similarity mapping, rigorous constraint encoding, and evaluation on production-grade annealing hardware, this study advances the application of QUBO-based optimization to real-world supply-chain planning. The results highlight the potential of hybrid quantum-classical approaches to deliver feasible, profitable, and diverse allocation strategies at industrial scale.
\end{abstract}\\

\section{Introduction}
Logistics and supply chain management tasks such as vehicle routing, scheduling, and inventory allocation are inherently combinatorial optimization problems. These problems require identifying the best solution from a finite set of possibilities. Unfortunately, many of them fall under the class of NP-hard problems, where the computational effort of classical exact solvers increases exponentially with the size of the problem instance, making them impractical for real-world large-scale systems \cite{ref1}.  

Due to these computational challenges, practitioners often rely on approximate solvers. Classical metaheuristics such as Genetic Algorithms (GA), Particle Swarm Optimization (PSO), and Ant Colony Optimization (ACO) dominate this space. These methods are well-suited to nonlinear and multi-modal cost functions commonly encountered in logistics, though they often require extensive parameter tuning and may converge to local optima, particularly in high-dimensional dynamic networks. Their scalability is also restricted by the curse of dimensionality. In response, researchers continue to explore improvements and hybrid strategies, including quantum-inspired variants \cite{ref1}.  

Quantum computing introduces novel paradigms for tackling optimization problems, exploiting superposition, entanglement, and tunneling. Popular approaches include Quantum Annealing (QA) and gate-based algorithms such as the Quantum Approximate Optimization Algorithm (QAOA) and Variational Quantum Eigensolver (VQE) \cite{ref2}. QA maps the problem into an Ising or QUBO model and searches the solution space using the quantum adiabatic theorem \cite{ref3}. Gate-based approaches like QAOA and VQE employ parameterized quantum circuits to variationally identify low-energy states \cite{ref2}. These techniques have been applied to benchmark problems such as TSP, Max-Cut, and Knapsack on quantum hardware. Furthermore, simulated quantum annealing—a classical analogue of QA—can inherit tunneling-like properties to efficiently explore the energy landscape and avoid local minima \cite{refk}.  

Despite their potential, quantum devices remain limited due to noise and restricted qubit counts. As a result, most practical applications employ hybrid quantum-classical schemes, where quantum subroutines complement classical solvers \cite{ref4}. In this study, we address the multiperiod SKU allocation problem in supply chain optimization via a Quantum inspired QUBO-based model implemented on Quanfluence's Time-Multiplexed Coherent Ising Machine (CIM), benchmarked against classical metaheuristics including GA, PSO, and ACO.

\section{Literature Review}
Optimization in supply chain management has been extensively studied using a wide spectrum of methods. The literature broadly falls into two categories: 
\textit{classical metaheuristics}, which provide scalable heuristic solutions on conventional hardware, and \textit{quantum or quantum-inspired approaches}, which aim to exploit tunneling and superposition effects to explore solution landscapes more efficiently. 
In this section, we first review the classical methods that remain the backbone of large-scale optimization, followed by recent advances in quantum optimization techniques relevant to logistics and supply chain problems.
\subsection{Classical Methods}
As discussed earlier, GA, PSO, and ACO represent the traditional workhorses of combinatorial optimization \cite{ref1}.  

GA, inspired by natural evolution, evolve a population of solutions through selection, crossover, and mutation. A notable application is presented in \cite{ref6}, where a real-coded GA was used to optimize logistics supply chain operations. The study demonstrated improvements in computational efficiency and provided integrators with tools to determine optimal operational strategies.  

PSO models the collective movement of agents, updating solutions based on both individual and swarm experiences \cite{ref7}. A significant application is described in \cite{ref8}, where a swarm intelligence framework for supply chain management enabled rapid convergence and improved flexibility in vehicle routing and resource allocation, outperforming exact solvers in scalability.  

ACO, motivated by the pheromone trail behavior of ants, incrementally improves candidate paths \cite{ref9}. In \cite{ref10}, ACO combined with TRIZ methodology was applied to optimize Printed Circuit Board (PCB) assembly operations, reducing assembly times and enhancing cost-effectiveness in supply chain contexts.

In summary, classical metaheuristics such as GA, PSO, and ACO remain the dominant approaches for large-scale combinatorial optimization in supply chains. Their flexibility and adaptability make them valuable in practice; however, they often struggle with parameter sensitivity, premature convergence, and scalability in high-dimensional multiperiod problems. These limitations motivate the exploration of alternative paradigms, particularly quantum and quantum-inspired approaches, which offer the potential to escape local optima and explore solution spaces more efficiently.

\subsection{Quantum Methods}
Quantum and quantum-inspired optimization techniques have emerged as promising approaches for tackling combinatorial problems in logistics and supply-chain management. Foundational studies established that many operational decision problems can be expressed as QUBO formulations, enabling direct use of annealing and Ising-based solvers \cite{lucas2014}. Early demonstrations with quantum annealers showed feasibility on canonical problems such as the Traveling Salesman Problem (TSP) and small-scale Vehicle Routing Problems (VRP), often in hybrid schemes combining quantum hardware with classical repair or decomposition heuristics \cite{neukart2017}. These works confirmed that annealing-based methods can explore large combinatorial landscapes efficiently, though scaling to realistic industrial constraints remains challenging.

Parallel advances in alternative hardware platforms, such as CIM, have demonstrated competitive performance on dense benchmark problems like MAX-CUT on a completely-connected graph, suggesting that non-gate-model physical solvers may also hold relevance for applied logistics tasks \cite{inagaki2016,mcmahon2016}. More recently, research has investigated quantum-enhanced feature representations, where variational circuits are used to embed classical data into high-dimensional Hilbert spaces for constructing similarity kernels \cite{havlicek2019}. While such embeddings have mainly been applied in machine-learning contexts, they hint at new opportunities for influencing combinatorial objectives by encoding correlations or redundancy penalties in quantum-derived form.

Despite these advances, most prior work has either focused on routing-type benchmarks or has used quantum components primarily as auxiliary preprocessing tools. Embedding realistic operational constraints—such as multi-period capacity limits, product diversity requirements, or risk penalties—directly into QUBO formulations remains underexplored. In particular, integrating quantum-derived similarity structure with strict feasibility enforcement in large-scale supply-chain optimization has received little attention, leaving a gap between proof-of-concept demonstrations and production-oriented formulations \cite{dwave2025,whitepaper2021}. This gap motivates the exploration of hybrid quantum-inspired approaches that not only minimize abstract energy functions but also yield solutions interpretable in terms of business-centric performance indicators.

\section{Quanfluence's Coherent Ising Machine}
Quanfluence Optimizer is a quantum inspired time-multiplexed coherent Ising machine. An opto-electronic gain dissipative loop evolves the decision variables while the interaction terms are implemented in a high-speed electronic circuit. Amplitude stabilization and solution diversification controls aid the machine to avoid getting 
trapped in local minima. The optimizer also features additional physics inspired algorithms that complement and support the main coherent Ising machine in exploring high quality solutions.
\section{ About the Data}
In this study, we analyze a supply chain dataset obtained from Kaggle \cite{amirmotefaker2021}, which provides a detailed record of operations within a consumer goods distribution network. The dataset includes information on 100 distinct SKUs, categorized into three primary product groups: haircare, skincare, and cosmetics. It offers a representative snapshot of real-world supply chain activity, capturing key aspects such as product movement, demand variability, and category-level differentiation, making it well-suited for research in supply chain analytics and operational optimization. 
Here’s what the dataset includes: each product has a unique SKU and belongs to one of the three categories, with skincare making up about 38\% of the mix, followed by haircare (32\%) and cosmetics (30\%). Stock availability and levels vary widely, from 1 to 100 units, showing how some products are always in demand while others sit on shelves.
The dataset dives into logistics too—lead times (how long it takes to get products ready) range from 1 to 30 days, and shipping times (how long delivery takes) go from 1 to 10 days. We see three shipping carriers (A, B, and C, with Carrier B leading at 43\%). Suppliers are spread across five major Indian cities—Mumbai, Kolkata, Chennai, Bangalore, and Delhi—with Mumbai and Kolkata being the busiest hubs. Manufacturing details, like production volumes and costs, give us insight into the production side, while quality checks or inspection results (Pass, Fail, or Pending) and defect rates (0.02\% to 4.94\%) highlight the challenges of maintaining product standards. Finally, transportation modes (Road, Rail, Air, Sea) and routes (A, B, C) round out the picture, showing how products move through the supply chain.

\subsection*{Key Features:}
\begin{itemize}
    \item \textbf{Unit Margin} – Profit per unit.
    \item \textbf{Demand} – Expected quantity over the period.
    \item \textbf{Total Cost} – Normalized cost for comparison.
    \item \textbf{Inventory Risk} – Risk of overstock/understock.
    \item \textbf{Defect Risk} – Likelihood of item defect.
    \item \textbf{Utilization \& Lead Time} – Operational metrics.
\end{itemize}

\section{ Feature Engineering and Data Expansion}

To enhance the analytical value of the supply chain dataset, we conducted a series of feature engineering steps aimed at improving interpretability and supporting advanced modeling. This involved deriving new variables and normalizing key metrics to better capture cost structures, profitability, operational efficiency, and supply chain risks.

\subsection{Derived Metrics}
Several new metrics were developed to provide deeper insights into profitability, production utilization, and operational bottlenecks:

\paragraph{TotalCost}
\textit{TotalCost} captures the comprehensive cost incurred in producing and delivering a product. It combines manufacturing costs, shipping expenses, and any additional operational overhead. This metric enables a holistic understanding of unit-level cost structure, which is essential for profitability analysis.

\[
\text{TotalCost} = C_m + C_s + C_o
\]
where:
\begin{itemize}
  \item $C_m$: Manufacturing cost
  \item $C_s$: Shipping cost
  \item $C_o$: Other operational costs (e.g., handling, packaging)
\end{itemize}

\paragraph{UnitMargin}
\textit{UnitMargin} measures the profit generated per unit sold by subtracting the total cost from the selling price. It helps identify products that are profitable versus those that may be operating at a loss.

\[
\text{UnitMargin} = P - \text{TotalCost}
\]
where $P$ is the product's selling price.

\paragraph{Utilization}
\textit{Utilization} reflects how efficiently production capacity is being used. It is calculated as the ratio of units sold to units that can be produced, providing insights into under- or over-utilized manufacturing resources.

\[
\text{Utilization} = \frac{Q_s}{Q_p}
\]
where:
\begin{itemize}
  \item $Q_s$: Quantity sold
  \item $Q_p$: Production volume or capacity
\end{itemize}

\paragraph{Overload}
\textit{Overload} is a binary indicator flagging whether demand exceeds production capacity. This metric is critical for detecting supply bottlenecks and ensuring the supply chain can meet customer demand.

\[
\text{Overload} =
\begin{cases}
1 & \text{if } Q_s > Q_p \\
0 & \text{otherwise}
\end{cases}
\]

\paragraph{InventoryRisk}
\textit{InventoryRisk} assesses the balance between stock levels and actual sales. A negative value indicates stockouts (inventory shortages), while a positive value reflects surplus inventory. This metric helps manage inventory efficiency and avoid both under- and over-stocking.

\[
\text{InventoryRisk} = \frac{I - Q_s}{Q_s}
\]
where:
\begin{itemize}
  \item $I$: Inventory level
\end{itemize}

\paragraph{LeadTimeRisk}
\textit{LeadTimeRisk} flags products with unusually long lead times, which can delay fulfillment and impact customer satisfaction. It marks products in the top 25\% of the lead time distribution.

\[
\text{LeadTimeRisk} =
\begin{cases}
1 & \text{if } L > L_{75} \\
0 & \text{otherwise}
\end{cases}
\]
where:
\begin{itemize}
  \item $L$: Lead time for the product
  \item $L_{75}$: 75th percentile of lead times across all products
\end{itemize}

\paragraph{DefectRisk}
\textit{DefectRisk} captures the quality risk by quantifying the defect rate for each product. If a product has known defects, its defect rate is used; otherwise, the risk is considered zero. This metric is vital for identifying quality issues in manufacturing.

\[
\text{DefectRisk} =
\begin{cases}
D & \text{if product fails inspection} \\
0 & \text{otherwise}
\end{cases}
\]
where $D$ is the defect rate (e.g., 0.0494 for 4.94\%).

\subsection{Normalization}

To ensure comparability across features and prepare the dataset for machine learning applications, continuous variables were normalized using min-max scaling:

\[
x_{\text{norm}} = \frac{x - \min(x)}{\max(x) - \min(x)}
\]

This transformation maps all values to the $[0, 1]$ range, ensuring no individual feature dominates due to scale differences.

\subsection{Data Expansion}
To support more comprehensive analysis and enable the development of robust models, we expanded the original dataset from 100 to 500 unique SKUs through the generation of synthetic data. This augmentation process was carefully designed to preserve the statistical properties and structural relationships observed in the original dataset, while introducing new, realistic records that do not replicate existing entries.

Key characteristics—such as the proportional distribution of product categories (haircare, skincare, and cosmetics), the relationship between pricing and revenue, and the variability in shipping and inventory-related features—were maintained to ensure consistency. Numerical variables, including cost components and lead times, were synthesized using a generative approach guided by the statistical properties and observed relationships in the original dataset. The synthetic data was designed to preserve realistic variability and internal consistency, mirroring key distributional and relational patterns without directly duplicating original records.


\section{ Formulations}
\subsection{Quantum QUBO formulation}
We formulate the multi‑period SKU allocation as a Quadratic Unconstrained Binary Optimization (QUBO) problem. 

\noindent\textbf{Symbol Definitions:}

\begin{tabular}{@{}ll@{}}
\toprule
Symbol & Meaning \\
\midrule
\( U_i \) & Unit margin for SKU \( i \) \\
\( D_i \) & Forecasted demand for SKU \( i \) \\
\( r_i \) & Unified risk for SKU \( i \) \\
\( S_{ij} \) & Similarity between SKUs \( i \) and \( j \) \\
\( C \) & Capacity limit per period \\
\( K \) & Target number of SKUs per period \\
\( s_b^{(t)} \) & Slack bit (capacity encoding) \\
\bottomrule
\end{tabular}
\bigskip
\\
\subsubsection{Capacity constraint via slack bits (implementation note)}

To enforce the per-period soft capacity constraint
\[
  \sum_{i}D_{i}\,x_{i}^{(t)} \le C
\]
we introduce \(B\) binary slack variables
\[
  s_{b}^{(t)} \in \{0,1\},\qquad b=0,\dots,B-1,
\]
which encode a nonnegative integer
\[
  S^{(t)} = \sum_{b=0}^{B-1}2^{b}\,s_{b}^{(t)}.
\]
In the implementation we use the penalty
\[
  H_{\rm cap}
  = \lambda_{c}
    \sum_{t=1}^{T}
    \Bigl(\sum_{i}D_{i}\,x_{i}^{(t)} - C + \sum_{b=0}^{B-1}2^{b}s_{b}^{(t)}\Bigr)^{2},
\]
(i.e. the slack sum enters with a \(+\) sign, matching the code). Expanding and removing the optimization-irrelevant constant yields the quadratic contributions used in the code:
\begin{align*}
  &\lambda_c\sum_{t}\Biggl[
    \sum_{i}D_{i}^{2}x_{i}^{(t)}
    + \sum_{b}2^{2b}s_{b}^{(t)}
    + 2\sum_{i<j}D_{i}D_{j}x_{i}^{(t)}x_{j}^{(t)}\\
  &\qquad\qquad
    + 2\sum_{i,b}2^{b}D_{i}x_{i}^{(t)}s_{b}^{(t)}
    + 2\sum_{b<b'}2^{b+b'}s_{b}^{(t)}s_{b'}^{(t)}
    -2C\sum_{i}D_{i}x_{i}^{(t)}
    -2C\sum_{b}2^{b}s_{b}^{(t)}
  \Biggr].
\end{align*}
Note: the implementation stores the constant term \(\lambda_c C^2\) explicitly in the QUBO dictionary at index \((0,0)\) to preserve numeric parity with our reference code; this constant does not affect the optimizer but is included for reproducibility.

\paragraph{Choice of \(B\).} In the code \(B\) is set by the variable \texttt{num\_slack\_bits}; in reported experiments we used \(B=13\) (i.e. \texttt{num\_slack\_bits=13}), which provided a good tradeoff between representational range and total variable count.

\subsubsection{ Overall Objective}

Our objective is to minimize a quadratic binary function:
\[
  \min_{\mathbf{z} \in \{0,1\}^{T(N+B)}}\; \mathbf{z}^\top Q\,\mathbf{z}
  \;=\;
  \sum_{t=1}^{T} \begin{bmatrix} x^{(t)} \\ s^{(t)} \end{bmatrix}^\top
  Q^{(t)} \begin{bmatrix} x^{(t)} \\ s^{(t)} \end{bmatrix},
\]
where each \( Q^{(t)} \in \mathbb{R}^{(N+B)\times(N+B)} \) encodes costs, penalties, and constraints for period~\( t \), and cross‑period coupling is set to zero in our experiments.

\subsubsection{ Margin (Profit) Term}
\[
  H_{\mathrm{margin}}
  = -\lambda_{m}\sum_{t=1}^{T}\sum_{i=1}^{N}
    (U_i \cdot D_i)\,x_{i}^{(t)}.
\]
Here \( U_i \) is the unit margin for SKU~\( i \), and \( D_i \) is the expected demand.

\subsubsection{ Quantum Similarity (Diversity) Term}
\[
  H_{\mathrm{sim}}
  = \lambda_{s}\sum_{t=1}^{T}\sum_{1\le i<j\le N}
    S_{ij}\,x_{i}^{(t)}\,x_{j}^{(t)},
\]
where \( S_{ij} = |\langle\psi(f_i)\mid\psi(f_j)\rangle|^2 \) is the quantum kernel similarity between features \( f_i \) and \( f_j \).

\subsubsection{ Risk \& Penalty Terms}
We add separate penalties for various risk metrics:
\[
\begin{aligned}
  H_{\mathrm{risk}}
    &= \lambda_{r}\sum_{t,i} r_i\,D_i\,x_{i}^{(t)},\\
  H_{\mathrm{inv}}
    &= \lambda_{\mathrm{inv}}\sum_{t,i} \mathrm{InventoryRisk}_i\,x_{i}^{(t)},\\
  H_{\mathrm{def}}
    &= \lambda_{\mathrm{def}}\sum_{t,i} \mathrm{DefectRisk}_i\,x_{i}^{(t)},
\end{aligned}
\]
where \( r_i \) denotes normalized unified risk for SKU~\( i \).

\subsubsection{ Capacity Constraint via Slack Bits}

To enforce the soft constraint \( \sum_i D_i x_i^{(t)} \le C \), we introduce a penalty:
\[
  H_{\mathrm{cap}}
  = \lambda_{c}\sum_{t=1}^{T}
    \left(\sum_i D_i\,x_{i}^{(t)} - C + \sum_{b=0}^{B-1}2^b\,s_{b}^{(t)}\right)^2.
\]
This introduces quadratic interactions among both decision variables and slack bits:
\begin{itemize}
  \item Demand² terms: \( \sum_{i,j} D_i D_j\,x_i x_j \)
  \item Slack² terms: \( \sum_{a,b} 2^{a+b}\,s_a s_b \)
  \item Cross terms: \( 2\sum_{i,b} D_i 2^b\,x_i s_b \)
  \item Linear terms: \( -2C \sum_i D_i x_i \), \( -2C \sum_b 2^b s_b \)
  \item Constant: \( +C^2 \)
\end{itemize}

\subsubsection{Cardinality (SKU‑Count) Constraint}

We enforce selecting exactly \(K\) SKUs in each period via
\[
  H_{\mathrm{card}}
  = \lambda_{k}
    \sum_{t=1}^{T}
    \Bigl(\sum_{i}x_{i}^{(t)} - K\Bigr)^{2}
  \;=\;\lambda_{k}
    \sum_{t=1}^{T}
    \Bigl[
      \sum_{i}x_{i}^{(t)}
      + 2\sum_{i<j}x_{i}^{(t)}x_{j}^{(t)}
      -2K\sum_{i}x_{i}^{(t)}
      + K^{2}
    \Bigr].
\]

Since \(K^2\) is a constant offset (irrelevant to the optimizer), we drop it. Thus the QUBO contributions for each period \(t\) are:

\[
\begin{aligned}
  &\text{Diagonal terms:}&
    &Q_{\,(t,i),(t,i)} \;\mathrel{+}= \;\lambda_{k} + \frac{\lambda_{\text{sku\_limit}}}{K}
      \;-\;2\lambda_{k}K, 
    \quad i=1,\dots,n,\\[6pt]
  &\text{Off‐diagonal terms:}&
    &Q_{\,(t,i),(t,j)} \;\mathrel{+}= \;2\lambda_{k},
    \quad 1 \le i < j \le n.
\end{aligned}
\]

This is implemented without auxiliary variables, by expanding the quadratic form into linear and pairwise terms. The additional term \(\lambda_{\text{sku\_limit}}/K\) provides fine-grained control over the SKU selection penalty independently of the cardinality enforcement strength \(\lambda_k\).

\subsubsection{ Top‑5 SKU Enforcement (Optional)}

To ensure high-margin SKUs are included, we penalize omission of the top‑5 items:
\[
  H_{\mathrm{top5}} = -\lambda_{\mathrm{top5}}\sum_{t=1}^{T}\sum_{i\in\mathcal{T}} x_{i}^{(t)},
\]
where \( \mathcal{T} \subset \{1,\dots,N\} \) is the set of indices of the top‑5 SKUs by \( U_i D_i \). The weight \( \lambda_{\mathrm{top5}} \) is chosen to overpower other terms.

\subsubsection{Combined QUBO Matrix}

For each period \(t\), let \(i,j=1,\dots,N+B\) index the decision and slack bits.  Then

\[
Q_{ij}^{(t)} \;=\;
\begin{cases}
\displaystyle
\underbrace{-\lambda_{m}U_iD_i}_{\text{margin}}
\;+\;
\underbrace{\lambda_{r}r_iD_i}_{\text{risk}}
\;+\;
\underbrace{\lambda_{\mathrm{inv}}\,\mathrm{InvRisk}_i}_{\substack{\text{inventory}\\\text{risk}}}
\;+\;
\underbrace{\lambda_{\mathrm{def}}\,\mathrm{DefectRisk}_i}_{\substack{\text{defect}\\\text{risk}}}
\;-\;
\underbrace{\lambda_{\mathrm{top5}}\mathbb{I}_{i\in\mathcal T}}_{\text{top-5 enforce}} \\[6pt]
\quad\displaystyle
+\;\underbrace{2\lambda_{k}}_{\substack{\text{cardinality}\\\text{off-diagonal}}}
\;+\;
\underbrace{\sum_{b=0}^{B-1}\Bigl(\lambda_c2^{2b}-2\lambda_cC\,2^b\Bigr)}_{\substack{\text{slack² \&}\\\text{linear terms}}}
\;, 
& i=j\le N, 
\\[12pt]
\displaystyle
\underbrace{\lambda_{s}S_{ij}}_{\substack{\text{similarity}\\\text{penalty}}}
\;+\;
\underbrace{\lambda_cD_iD_j}_{\substack{\text{capacity}\\\text{cross-term}}}
\;+\;
\underbrace{2\lambda_{k}}_{\text{cardinality}}
\;, 
& i\neq j \le N,
\\[8pt]
\displaystyle
\underbrace{-2\lambda_cD_i2^b}_{\substack{\text{demand–slack}\\\text{cross-term}}}
\;, 
& i\le N,\; j = N + b,
\\[8pt]
\displaystyle
\underbrace{2\lambda_c2^{a+b}}_{\substack{\text{slack–slack}\\\text{cross-term}}}
\;, 
& i = N + a,\; j = N + b,
\\[4pt]
0, & \text{otherwise.}
\end{cases}
\]

The full problem is:
\[
  \min_{\mathbf{z} \in \{0,1\}^{T(N+B)}}\; \mathbf{z}^\top Q\,\mathbf{z},
  \quad \mathbf{z} = [x^{(1)}, s^{(1)}, \dots, x^{(T)}, s^{(T)}]^\top.
\]

\bigskip
\noindent\textbf{Hyperparameter Mapping:}

\begin{tabular}{@{}ll@{}}
\toprule
Theory symbol & Code variable \\
\midrule
$\lambda_{m}$ (margin)            & \texttt{margin\_weight} \\
$\lambda_{s}$ (similarity)        & \texttt{similarity\_weight} \\
$\lambda_{r}$ (unified risk)      & \texttt{penalty\_risk} \\
$\lambda_{\mathrm{inv}}$ (inv. risk) & \texttt{penalty\_inventory} \\
$\lambda_{\mathrm{def}}$ (defect risk) & \texttt{penalty\_defect} \\
$\lambda_{c}$ (capacity)          & \texttt{penalty\_capacity} \\
$\lambda_{k}$ (cardinality)       & \texttt{penalty\_k} \\
$\lambda_{\text{sku\_limit}}$ (SKU limit penalty) & \texttt{penalty\_sku\_limit} \\  

$\lambda_{\mathrm{top5}}$ (top‑5 enforcement) & \texttt{enforce\_P} \\
$K$ (SKU limit per period)        & \texttt{max\_skus\_per\_period} \\
$C$ (capacity per period)         & \texttt{capacity\_per\_period} \\
\bottomrule
\end{tabular}

\bigskip
\noindent\textbf{Parameter tuning:}  
All penalty weights \( \lambda \) are selected through empirical sensitivity analysis. We systematically vary the penalties and observe their effect on profit, diversity, risk, and constraint satisfaction across held-out scenarios. This process helps tune the parameters to achieve a balanced and feasible solution, rather than relying on formal cross-validation.

\begin{itemize}
\item[-] The implementation uses zero-based indexing for periods and variables (periods \(t=0,\dots,T-1\)); the math above uses \(t=1,\dots,T\) for notational clarity.  
\item[-] The QUBO is stored as a sparse dictionary of symmetric pairs with keys \((i,j)\) where keys are sorted (this preserves symmetry). The constant offset \(\lambda_c C^2\) is stored at key \((0,0)\) in the implementation.
\end{itemize}

\subsection{Classical PSO Formulation}

We solve the same multi-period SKU allocation problem using \textit{Particle Swarm Optimization}, where each particle encodes a binary allocation vector over all periods.

\subsubsection*{Symbol Definitions}

\begin{table}[H]
\centering
\begin{tabular}{ll}
\hline
\textbf{Symbol} & \textbf{Meaning} \\
\hline
$\mathbf{x}_p^{(t)}$ & Position vector of particle $p$ at iteration $t$ (SKU selection decisions) \\
$\mathbf{v}_p^{(t)}$ & Velocity vector of particle $p$ at iteration $t$ \\
$\mathbf{pbest}_p$ & Best position found by particle $p$ so far \\
$\mathbf{gbest}$ & Global best position among all particles \\
$N$ & Number of SKUs \\
$T$ & Number of time periods \\
$C$ & Capacity per period \\
$K$ & Maximum SKUs per period \\
$U_i$ & Unit margin for SKU $i$ \\
$D_i$ & Forecasted demand for SKU $i$ \\
$r_i$ & Normalized risk score for SKU $i$ \\
$S_{ij}$ & Cosine similarity between SKU $i$ and $j$ in PCA space \\
$\mathcal{T}$ & Index set of top-5 SKUs by $U_i D_i$ \\
$\lambda_m, \lambda_s, \lambda_r, \lambda_c, \lambda_k,$ & Same penalty weights as in QUBO \\
$\lambda_{\mathrm{inv}}, \lambda_{\mathrm{def}}, \lambda_{\mathrm{top5}}$ & \\
\hline
\end{tabular}
\end{table}

\subsubsection*{PSO Dynamics}

\paragraph{Velocity update:}
\begin{equation}
\mathbf{v}_p^{(t+1)}
= \omega\,\mathbf{v}_p^{(t)}
+ c_1\,\mathbf{r}_1^{(t)} \circ (\mathbf{pbest}_p - \mathbf{x}_p^{(t)})
+ c_2\,\mathbf{r}_2^{(t)} \circ (\mathbf{gbest} - \mathbf{x}_p^{(t)})
\end{equation}
where $\omega$ = inertia weight, $c_1, c_2$ = cognitive/social coefficients, $\mathbf{r}_1, \mathbf{r}_2 \sim U(0,1)^d$, and $\circ$ = elementwise multiplication.

\paragraph{Position update:}
\begin{equation}
\mathbf{x}_p^{(t+1)} = \mathrm{clip}\big(\mathbf{x}_p^{(t)} + \mathbf{v}_p^{(t+1)},\,0,\,1\big)
\end{equation}
Binary mapping:
\begin{equation}
x_{p,j}^{(t+1)} =
\begin{cases}
1, & \sigma(v_{p,j}^{(t+1)}) > u, \quad u \sim U(0,1) \\
0, & \text{otherwise}
\end{cases}
\quad
\sigma(z) = \frac{1}{1+e^{-z}}
\end{equation}

Top-5 enforcement:
\begin{equation}
x_{p,j}^{(t+1)} = 1, \quad \forall j \in \mathcal{T},\; \forall t
\end{equation}

\subsubsection*{Fitness Function}

The objective function is defined as:
\begin{align}
F(\mathbf{x}) &=
\underbrace{\lambda_s \sum_{t=1}^T \sum_{i<j} S_{ij} x_i^{(t)} x_j^{(t)}}_{\text{similarity penalty}}
-
\underbrace{\lambda_m \sum_{t=1}^T \sum_{i=1}^N (U_i D_i)\,x_i^{(t)}}_{\text{profit term}} \nonumber \\
&\quad + \underbrace{\lambda_r \sum_{t=1}^T \sum_{i=1}^N r_i D_i\,x_i^{(t)}}_{\text{risk penalty}}
+ \underbrace{\lambda_{\mathrm{inv}} \sum_{t=1}^T \sum_{i=1}^N \mathrm{InvRisk}_i\,x_i^{(t)}}_{\text{inventory penalty}} \nonumber \\
&\quad + \underbrace{\lambda_{\mathrm{def}} \sum_{t=1}^T \sum_{i=1}^N \mathrm{DefectRisk}_i\,x_i^{(t)}}_{\text{defect penalty}}
+ \underbrace{\lambda_k \sum_{t=1}^T \left(\sum_{i=1}^N x_i^{(t)} - K\right)^2}_{\text{SKU count constraint}} \nonumber \\
&\quad + \underbrace{\lambda_c \sum_{t=1}^T \max\left(0, \sum_{i=1}^N D_i x_i^{(t)} - C\right)^6}_{\text{capacity constraint (6th order)}}
- \underbrace{\lambda_{\mathrm{top5}} \sum_{t=1}^T \sum_{i \in \mathcal{T}} x_i^{(t)}}_{\text{top-5 enforcement}}
\end{align}

\subsubsection*{Optimization Problem}
\begin{align}
&\min_{\mathbf{x} \in \{0,1\}^{TN}} F(\mathbf{x}) \\
&\text{s.t.} \quad \mathbf{x} \ \text{evolves according to the PSO velocity--position update rules} \nonumber \\
&\quad\quad \text{and fixed selection of top-5 SKUs in all periods.} \nonumber
\end{align}

\subsection{Genetic Algorithm Formulation}

We solve the multi-period SKU allocation problem using a \textit{Genetic Algorithm}, where each individual in the population represents a binary allocation vector encoding SKU selections across all periods.

\subsubsection*{Symbol Definitions}

\begin{table}[H]
\centering
\begin{tabular}{ll}
\hline
\textbf{Symbol} & \textbf{Meaning} \\
\hline
$\mathbf{x}_k$ & Binary allocation vector for individual $k$ (SKU selection decisions) \\
$N$ & Number of SKUs \\
$T$ & Number of time periods \\
$C$ & Capacity per period (28392 units) \\
$K$ & Maximum SKUs per period (50) \\
$U_i$ & Unit margin for SKU $i$ \\
$D_i$ & Forecasted demand for SKU $i$ \\
$r_i$ & Normalized risk score for SKU $i$ \\
$S_{ij}$ & Cosine similarity between SKU $i$ and $j$ in PCA space \\
$\mathcal{T}$ & Index set of top-5 SKUs by $U_i D_i$ \\
$\lambda_s$ & Weight for similarity penalty (3.0) \\
$\lambda_m$ & Weight for profit term (0.02) \\
$\lambda_r$ & Weight for risk penalty (0.02) \\
$\lambda_{\mathrm{inv}}$ & Weight for inventory risk penalty (50) \\
$\lambda_{\mathrm{def}}$ & Weight for defect risk penalty (50) \\
$\lambda_k$ & Weight for quadratic SKU count penalty (1000) \\
$\lambda_{sku}$ & Weight for squared SKU count excess penalty (5000) \\
$\lambda_c$ & Base weight for capacity penalty (5000, scaled by $10^7$) \\
$\lambda_{\mathrm{top5}}$ & Large penalty for top-5 SKU enforcement ($10^9 \cdot \max(|U_i D_i|)$) \\
\hline
\end{tabular}
\end{table}

\subsubsection*{GA Dynamics}

The GA evolves a population of size 50 over 100 generations, with a crossover rate of 0.8 and a mutation rate of 0.1. The process includes the following steps:

\paragraph{Initialization:}
The population is initialized as a set of binary vectors $\{\mathbf{x}_k \mid k = 1, \dots, 50\}$, where each $\mathbf{x}_k \in \{0,1\}^{T \cdot N}$. For each individual, positions corresponding to the top-5 SKUs (indices in $\mathcal{T}$) are set to 1 for all periods to enforce their selection.

\paragraph{Selection:}
Select the top half of individuals (25) with the lowest fitness values to form the parent set. This elitist selection ensures the best solutions are preserved.

\paragraph{Crossover:}
For each offspring, with probability 0.8, perform single-point crossover between two parents. A random crossover point is chosen, and the offspring inherits the first segment from one parent and the second from the other. Top-5 SKU positions are fixed at 1 post-crossover:
\begin{equation}
x_{k,j} = 1, \quad \forall j \in \mathcal{T}, \forall t
\end{equation}

\paragraph{Mutation:}
For each offspring, each position $j$ (except those in $\mathcal{T}$) is flipped (0 to 1 or 1 to 0) with probability 0.1, preserving top-5 SKU selections:
\begin{equation}
x_{k,j} = 1 - x_{k,j}, \quad \text{if } \text{rand}() < 0.1 \text{ and } j \notin \mathcal{T}
\end{equation}

\paragraph{Final Adjustment:}
After evolution, the best solution is adjusted to ensure capacity compliance by deselecting non-top-5 SKUs with highest demand in periods where $\sum_{i=1}^N D_i x_{t,i} > C$, until the constraint is satisfied.

\subsubsection*{Fitness Function}

The fitness function to minimize is defined as:
\begin{align}
F(\mathbf{x}) &=
\underbrace{\lambda_s \sum_{t=1}^T \sum_{i=1}^{N-1} \sum_{j=i+1}^N S_{ij} x_{t,i} x_{t,j}}_{\text{similarity penalty}}
-
\underbrace{\lambda_m \sum_{t=1}^T \sum_{i=1}^N (U_i D_i) x_{t,i}}_{\text{profit term}} \nonumber \\
&\quad + \underbrace{\lambda_r \sum_{t=1}^T \sum_{i=1}^N r_i D_i x_{t,i}}_{\text{risk penalty}}
+ \underbrace{\lambda_{\mathrm{inv}} \sum_{t=1}^T \sum_{i=1}^N \mathrm{InvRisk}_i x_{t,i}}_{\text{inventory penalty}} \nonumber \\
&\quad + \underbrace{\lambda_{\mathrm{def}} \sum_{t=1}^T \sum_{i=1}^N \mathrm{DefectRisk}_i x_{t,i}}_{\text{defect penalty}}
+ \underbrace{\lambda_k \sum_{t=1}^T \left( \left( \sum_{i=1}^N x_{t,i} \right)^2 - 2K \sum_{i=1}^N x_{t,i} \right)}_{\text{quadratic SKU count penalty}} \nonumber \\
&\quad + \underbrace{\lambda_{sku} \sum_{t=1}^T \left( \max\left(0, \sum_{i=1}^N x_{t,i} - K\right) \right)^2}_{\text{SKU count excess penalty}}
+ \underbrace{\lambda_c \cdot 10^7 \sum_{t=1}^T \left( \max\left(0, \sum_{i=1}^N D_i x_{t,i} - C\right) \right)^6}_{\text{capacity penalty (6th order)}} \nonumber \\
&\quad + \underbrace{\lambda_{\mathrm{top5}} \sum_{t=1}^T \sum_{i \in \mathcal{T}} (1 - x_{t,i})}_{\text{top-5 enforcement penalty}}
\end{align}

\subsubsection*{Optimization Problem}

\begin{align}
&\min_{\mathbf{x} \in \{0,1\}^{T \cdot N}} F(\mathbf{x}) \\
&\text{s.t.} \quad x_{t,i} = 1, \quad \forall t = 1,\dots,T, \quad \forall i \in \mathcal{T} \quad (\text{top-5 SKU enforcement}) \nonumber \\
&\quad\quad \sum_{i=1}^N D_i x_{t,i} \leq C, \quad \forall t = 1,\dots,T \quad (\text{capacity constraint, enforced softly and adjusted post-GA}) \nonumber \\
&\quad\quad \sum_{i=1}^N x_{t,i} \leq K, \quad \forall t = 1,\dots,T \quad (\text{SKU limit, enforced softly}) \nonumber \\
&\quad\quad \text{Evolution follows GA selection, crossover, and mutation rules} \nonumber
\end{align}

\subsection{Ant Colony Optimization Formulation}

We solve the multi-period SKU allocation problem using \textit{Ant Colony Optimization}, where each ant constructs a binary allocation vector encoding SKU selections across all periods, guided by pheromone trails and heuristic information.

\subsubsection*{Symbol Definitions}

\begin{table}[H]
\centering
\begin{tabular}{ll}
\hline
\textbf{Symbol} & \textbf{Meaning} \\
\hline
$\mathbf{x}_k$ & Binary allocation vector for ant $k$ (SKU selection decisions) \\
$\tau_{j,c}$ & Pheromone level for variable $j$ (SKU-period pair) and choice $c \in \{0, 1\}$ \\
$\eta_j$ & Heuristic information for variable $j$ (based on normalized profit) \\
$N$ & Number of SKUs \\
$T$ & Number of time periods \\
$C$ & Capacity per period (28392 units) \\
$K$ & Maximum SKUs per period (50) \\
$U_i$ & Unit margin for SKU $i$ \\
$D_i$ & Forecasted demand for SKU $i$ \\
$r_i$ & Normalized risk score for SKU $i$ \\
$S_{ij}$ & Cosine similarity between SKU $i$ and $j$ in PCA space \\
$\mathcal{T}$ & Index set of top-5 SKUs by $U_i D_i$ \\
$\lambda_s$ & Weight for similarity penalty (3.0) \\
$\lambda_m$ & Weight for profit term (0.02) \\
$\lambda_r$ & Weight for risk penalty (0.02) \\
$\lambda_{\mathrm{inv}}$ & Weight for inventory risk penalty (50) \\
$\lambda_{\mathrm{def}}$ & Weight for defect risk penalty (50) \\
$\lambda_k$ & Weight for quadratic SKU count penalty (1000) \\
$\lambda_{sku}$ & Weight for squared SKU count excess penalty (5000) \\
$\lambda_c$ & Base weight for capacity penalty (5000, scaled by $10^7$) \\
$\lambda_{\mathrm{top5}}$ & Large penalty for top-5 SKU enforcement ($10^9 \cdot \max(|U_i D_i|)$) \\
$\alpha$ & Pheromone influence parameter (1.0) \\
$\beta$ & Heuristic influence parameter (2.0) \\
$\rho$ & Pheromone evaporation rate (0.5) \\
$q$ & Pheromone deposit constant (100.0) \\
\hline
\end{tabular}
\end{table}

\subsubsection*{ACO Dynamics}

The ACO algorithm runs with 50 ants over 100 iterations, using pheromone trails and a profit-based heuristic to guide solution construction. The process includes the following steps:

\paragraph{Pheromone Initialization:}
Pheromone trails are initialized uniformly for each variable $j \in \{1, \dots, T \cdot N\}$ and choice $c \in \{0, 1\}$:
\begin{equation}
\tau_{j,c} = 1.0, \quad \forall j, c
\end{equation}

\paragraph{Solution Construction:}
Each ant $k$ constructs a solution $\mathbf{x}_k \in \{0,1\}^{T \cdot N}$. For each variable $j$ (corresponding to SKU $i$ in period $t$), the probability of setting $x_{k,j} = 1$ is:
\begin{equation}
P(x_{k,j} = 1) = \frac{\tau_{j,1}^\alpha \cdot \eta_j^\beta}{\tau_{j,0}^\alpha + \tau_{j,1}^\alpha \cdot \eta_j^\beta}
\end{equation}
where $\eta_j = \frac{U_i D_i}{\max_i (U_i D_i)}$ is the heuristic information for variable $j$ (repeated across periods), and $x_{k,j} = 1$ is chosen if a random number $u \sim U(0,1) < P(x_{k,j} = 1)$. For top-5 SKUs:
\begin{equation}
x_{k,j} = 1, \quad \forall j \in \mathcal{T}, \forall t
\end{equation}
After construction, capacity is enforced by deselecting non-top-5 SKUs with highest demand in periods where $\sum_{i=1}^N D_i x_{t,i} > C$, until the constraint is satisfied.

\paragraph{Pheromone Update:}
Pheromones evaporate and are updated based on the best solution in the iteration:
\begin{equation}
\tau_{j,c} \gets (1 - \rho) \tau_{j,c} + \frac{q}{1 + F(\mathbf{x}_{\text{best}})} \cdot \mathbb{I}(x_{\text{best},j} = c)
\end{equation}
where $\mathbf{x}_{\text{best}}$ is the best solution found in the iteration, $F(\mathbf{x}_{\text{best}})$ is its fitness, and $\mathbb{I}$ is the indicator function.

\paragraph{Final Adjustment:}
After iterations, the best solution is adjusted to ensure capacity compliance by deselecting non-top-5 SKUs with highest demand in periods where $\sum_{i=1}^N D_i x_{t,i} > C$, until the constraint is satisfied.

\subsubsection*{Fitness Function}

The fitness function to minimize is defined as:
\begin{align}
F(\mathbf{x}) &=
\underbrace{\lambda_s \sum_{t=1}^T \sum_{i=1}^{N-1} \sum_{j=i+1}^N S_{ij} x_{t,i} x_{t,j}}_{\text{similarity penalty}}
-
\underbrace{\lambda_m \sum_{t=1}^T \sum_{i=1}^N (U_i D_i) x_{t,i}}_{\text{profit term}} \nonumber \\
&\quad + \underbrace{\lambda_r \sum_{t=1}^T \sum_{i=1}^N r_i D_i x_{t,i}}_{\text{risk penalty}}
+ \underbrace{\lambda_{\mathrm{inv}} \sum_{t=1}^T \sum_{i=1}^N \mathrm{InvRisk}_i x_{t,i}}_{\text{inventory penalty}} \nonumber \\
&\quad + \underbrace{\lambda_{\mathrm{def}} \sum_{t=1}^T \sum_{i=1}^N \mathrm{DefectRisk}_i x_{t,i}}_{\text{defect penalty}}
+ \underbrace{\lambda_k \sum_{t=1}^T \left( \left( \sum_{i=1}^N x_{t,i} \right)^2 - 2K \sum_{i=1}^N x_{t,i} \right)}_{\text{quadratic SKU count penalty}} \nonumber \\
&\quad + \underbrace{\lambda_{sku} \sum_{t=1}^T \left( \max\left(0, \sum_{i=1}^N x_{t,i} - K\right) \right)^2}_{\text{SKU count excess penalty}}
+ \underbrace{\lambda_c \cdot 10^7 \sum_{t=1}^T \left( \max\left(0, \sum_{i=1}^N D_i x_{t,i} - C\right) \right)^6}_{\text{capacity penalty (6th order)}} \nonumber \\
&\quad + \underbrace{\lambda_{\mathrm{top5}} \sum_{t=1}^T \sum_{i \in \mathcal{T}} (1 - x_{t,i})}_{\text{top-5 enforcement penalty}}
\end{align}

\subsubsection*{Optimization Problem}

\begin{align}
&\min_{\mathbf{x} \in \{0,1\}^{T \cdot N}} F(\mathbf{x}) \\
&\text{s.t.} \quad x_{t,i} = 1, \quad \forall t = 1,\dots,T, \quad \forall i \in \mathcal{T} \quad (\text{top-5 SKU enforcement}) \nonumber \\
&\quad\quad \sum_{i=1}^N D_i x_{t,i} \leq C, \quad \forall t = 1,\dots,T \quad (\text{capacity constraint, enforced softly and adjusted post-ACO}) \nonumber \\
&\quad\quad \sum_{i=1}^N x_{t,i} \leq K, \quad \forall t = 1,\dots,T \quad (\text{SKU limit, enforced softly}) \nonumber \\
&\quad\quad \text{Solutions constructed and pheromones updated per ACO rules} \nonumber
\end{align}

\section{ Methodology}

In this section, we describe in detail the workflow used to implement and evaluate our multi‑period SKU allocation model. We structure the methodology into six stages: data preparation, feature engineering, quantum embedding and similarity, QUBO construction, solution via simulated quantum annealing, and post‑processing analysis.  

\subsection{Data Preparation}
We begin by loading a cleaned, feature‐engineered dataset of SKU attributes.  Each row corresponds to one SKU and includes the following raw fields: demand ($D_i$), unit margin, defect risk, inventory risk, utilization, lead time, and total cost.  Missing values in any of these fields are removed prior to further processing.  The final dataset contains $500$ SKUs, each observed over $T=8$ consecutive planning periods.

\subsection{Feature Engineering and Normalization}
To capture multiple performance dimensions on a common scale, we normalize each raw attribute to zero mean and unit variance using standard score normalization \cite{normalization_review}. Specifically, for each feature column $f$, we compute
\[
\tilde f_i = \frac{f_i - \mu(f)}{\sigma(f)}\,,
\]
where $\mu(f)$ and $\sigma(f)$ are the empirical mean and standard deviation.  We then apply Principal Component Analysis (PCA) \cite{pca_ml} to reduce the five normalized features—unit–cost ratio, total cost, inventory risk, utilization, and lead time—down to $n_{\mathrm{q}}=5$ principal components.  This yields a compact, orthogonal embedding $\mathbf{x}_i\in\mathbb{R}^{5}$ for each SKU, which feeds directly into the quantum similarity circuit.

\subsection{Model Architecture}

\begin{figure}
    \centering
    \includegraphics[width=1.0\textwidth]{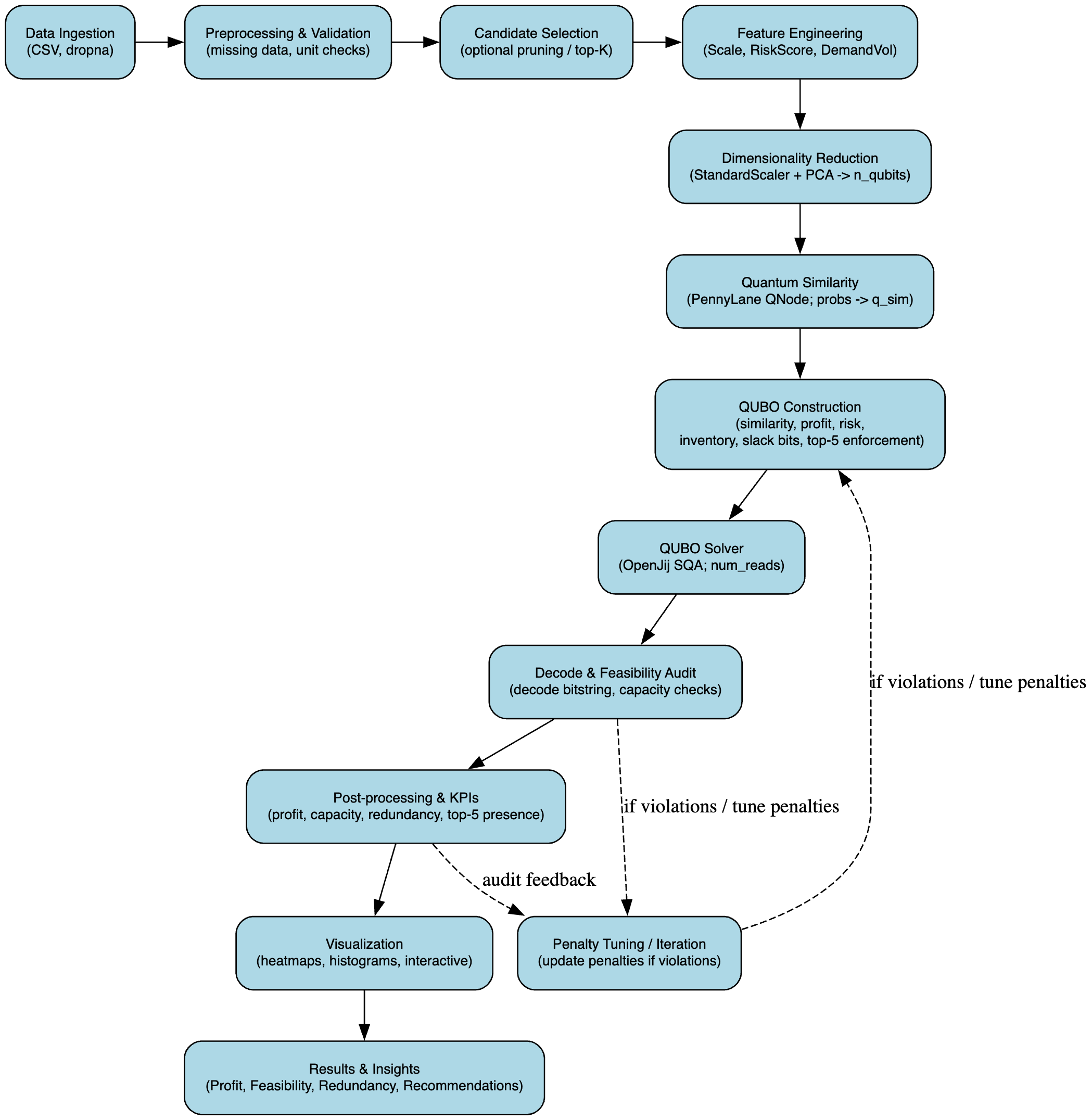}
    \caption{End-to-end pipeline of the quantum similarity-driven QUBO framework}
    \label{fig:model_arch}
\end{figure}

The proposed quantum-enhanced optimization framework operates through a systematic pipeline that transforms supply chain allocation problems into quantum-compatible formulations. Figure \ref{fig:model_arch} illustrates the complete architecture.

The process begins with raw data ingestion from CSV files or database connections, followed by preprocessing steps that handle missing values and validate measurement units. We then apply optional candidate selection through pruning or top-K filtering to reduce problem dimensionality. The core innovation lies in our feature engineering stage, where we compute three key metrics: Scale (normalized demand), RiskScore (supply uncertainty), and DemandVol (demand volatility). These features undergo standardization and PCA-based dimensionality reduction before being encoded into n-qubits.

The quantum similarity module leverages PennyLane's QNode\footnote{https://docs.pennylane.ai/en/stable/code/api/pennylane.QNode.html} architecture to compute pairwise similarities between SKUs. These similarity scores, along with profitability metrics, risk factors, inventory constraints, and slack variables, are fed into the QUBO construction phase. The resulting QUBO problem encodes business objectives like maximizing profit while satisfying capacity limits and enforcing top-5 SKU presence rules.

We solve the formulated QUBO using two quantum approaches: Quantfluence's CIM and OpenJij's simulated quantum annealing solver. Both solvers perform multiple reads to account for solution variability inherent in quantum sampling. The decoded bitstrings undergo feasibility audits to verify capacity constraints. Post-processing calculates KPIs including profit, capacity utilization, redundancy, and top-5 SKU presence. If violations occur, we enter an iterative penalty tuning loop that adjusts constraint weights and re-solves until feasibility is achieved. Finally, results are visualized through heatmaps and histograms, providing actionable insights for supply chain managers.

\subsection{Quantum Embedding and Similarity Matrix}

We employ the PennyLane \cite{pennylane} quantum computing framework with the “lightning.qubit” device \cite{lightning} to compute a fidelity–based similarity between every pair of PCA embeddings.  For two SKUs with embeddings $\mathbf{x}_i$ and $\mathbf{x}_j$, we prepare the quantum state
\[
\lvert\psi(\mathbf{x})\rangle = \bigotimes_{k=1}^{5} R_{X}(x_k)\lvert 0\rangle,
\]
apply $R_X$ rotations with opposite angles, and measure the probability of the all‑zero outcome. Following the quantum kernel approach \cite{havlicek,schuld}, for each pair $(i,j)$ we run
\[
p_{ij} = \bigl\lvert\langle 0\ldots0\rvert R_X(\mathbf{x}_i)\,R_X(-\mathbf{x}_j)\rvert 0\ldots0\rangle\bigr\rvert^2,
\]
and set $S_{ij}=S_{ji}=p_{ij}$.  We fill the diagonal with $S_{ii}=1$.  This yields a symmetric similarity matrix $S\in[0,1]^{N\times N}$.

As shown in Figure~\ref{fig:rx_embedding}, the quantum circuit demonstrates how penny lane generates the quantum similarity matrix to calculate quantum similarity: 

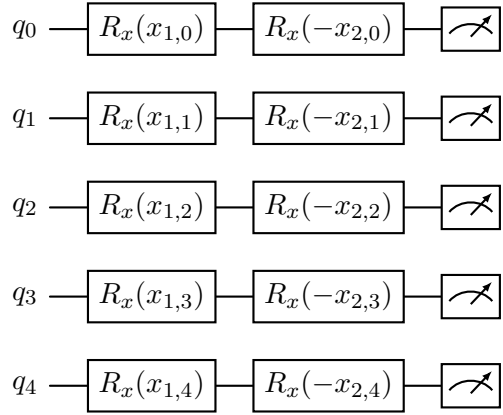
\begin{figure}[ht]
\centering
\begin{quantikz}
\lstick{$q_0$} & \gate{R_x(x_{1,0})} & \gate{R_x(-x_{2,0})} & \meter{} \\
\lstick{$q_1$} & \gate{R_x(x_{1,1})} & \gate{R_x(-x_{2,1})} & \meter{} \\
\lstick{$q_2$} & \gate{R_x(x_{1,2})} & \gate{R_x(-x_{2,2})} & \meter{} \\
\lstick{$q_3$} & \gate{R_x(x_{1,3})} & \gate{R_x(-x_{2,3})} & \meter{} \\
\lstick{$q_4$} & \gate{R_x(x_{1,4})} & \gate{R_x(-x_{2,4})} & \meter{} 
\end{quantikz}
\caption{RX-embedding used to compute the pairwise similarity between two SKU feature vectors \(x_1\) and \(x_2\). Each wire represents a qubit; for qubit \(i\) we apply $R_x(x_{1,i})$ followed by $R_x(-x_{2,i})$ and then measure the output probability distribution. The entry \(p_{00\ldots0}\) of the resulting probability vector is used as the similarity measure in the QUBO.}
\label{fig:rx_embedding}
\end{figure}

\subsection{QUBO Construction}
We map the SKU‐selection problem into a single large QUBO defined over $T(N + B)$ binary variables, where $B=13$ slack bits per period to encode capacity constraints up to $C=28{,}392$.  The overall Hamiltonian is
\[
H = H_{\rm margin} + H_{\rm sim} + H_{\rm risk} + H_{\rm inv} + H_{\rm def} + H_{\rm cap} + H_{\rm card}.
\]
Each term is translated into a weighted sum of quadratic and linear coefficients in a Python dictionary \texttt{Q}:  
\begin{itemize}
  \item \emph{Margin term} ($\lambda_m=0.02$): $-\sum_i(\text{unit\_margin}_i\times D_i)x_i^{(t)}$.  
  \item \emph{Similarity term} ($\lambda_s=1.0$): $\sum_{i<j}S_{ij}\,x_i^{(t)}x_j^{(t)}$.  
  \item \emph{Risk terms} ($\lambda_r=0.02$, $\lambda_{\rm inv}=50$, $\lambda_{\rm def}=50$): linear penalties $r_i D_i\,x_i$, inventory\,/\,defect risk as separate linear costs.  
  \item \emph{Capacity term} ($\lambda_c=5000$): squared constraint 
    \[
      \bigl(\sum_i D_i x_i^{(t)} - C - \sum_{b}2^b s_b^{(t)}\bigr)^2,
    \]
    fully expanded into demand–demand, slack–slack, demand–slack, linear, and constant contributions.  
  \item \emph{Cardinality term} ($\lambda_k=1000$): softly enforce $\sum_i x_i^{(t)}\le K=50$ via $(\sum_i x_i - K)^2\,. $  
\end{itemize}

All coefficients are assembled in nested loops over periods, SKU indices, and slack bits, and stored in \texttt{Q[(u,v)]}.


\subsection{QUBO Solving via Simulated Quantum Annealing}
The complete QUBO dictionary is passed to OpenJij's \cite{openjij,openjij_framework} SQASampler, an open-source QUBO solver, requesting 500 reads per problem instance. Alongside, we also leverage Quanfluence coherent Ising machine \cite{quanfluence} to solve the QUBO formulation. We call
\begin{verbatim}
    response = sampler.sample_qubo(Q, num_reads=500)
\end{verbatim}
and extract the lowest–energy sample. Slack bits automatically satisfy capacity constraints up to tolerances determined by $\lambda_c$.

\subsection{Post‑Processing and Analysis}
From the returned bitstring, we identify selected SKUs in each period.  We then:
\begin{enumerate}
  \item Compute total demand, profit, and cost across all periods.
  \item Check for any period exceeding capacity and report violations.
  \item Verify presence of the top‑5 most profitable SKUs in every period by imposing an additional negative bias in \texttt{Q}.
  \item Plot histograms of utilization and time series of period demands against capacity.
  \item Generate interactive heatmaps of the full similarity matrix and the submatrix of selected SKUs.
  \item Calculate a redundancy score among selected SKUs by averaging pairwise similarity multiplied by normalized margin.
\end{enumerate}

This six‐step methodology—from raw data to final analysis—ensures a transparent, reproducible pipeline and forms the basis for the results presented in Section~\ref{results}.

\section{Results Comparison}
\label{results}

In this section, we present the outcomes of our optimization experiments across six approaches: Quanfluence Quantum QUBO, Quanfluence Cosine QUBO, Cosine QUBO, Particle Swarm Optimization, Genetic Algorithm, and Ant Colony Optimization. The aim of this comparison is to evaluate not only profit, cost, and SKU selection, but also the ability of each method to maintain the inclusion of the top-5 most profitable SKUs across all planning periods. Alongside numerical summaries, we provide visualizations of maximum capacity usage and SKU similarity to better illustrate allocation patterns.

\begin{table}[h!]
\centering
\caption{Comparison of Optimization Results Across Methods}
\label{tab:results-comparison}
\begin{adjustbox}{width=\textwidth}
\begin{tabular}{lcccc}
\toprule
\textbf{Method} & \textbf{Total SKUs Selected} & \textbf{Units Selected} & \textbf{Net Profit} & \textbf{Total Cost} \\ 
\midrule
Quanfluence Quantum QUBO   & 288 & 226,813 & \$12,752,661.09 & \$187,182.28 \\
Quanfluence Cosine QUBO  & 283 & 221,611 & \$12,210,262.79 & \$172672.61 \\
SQA Quantum QUBO   & 191 & 198,907 & \$11,894,758.25  & \$157,290.07  \\
SQA Cosine QUBO    & 180  & 194,041  & \$11,700,278.44  & \$154,947.81  \\
PSO           & 83  & 49,755  & \$4,079,022.82  & \$25,344.50  \\
GA            & 271 & 225,357 & \$11,304,904.34 & \$168,586.61 \\
ACO           & 277 & 204,766 & \$10,518,980.76 & \$160,129.82 \\ 
\bottomrule
\end{tabular}
\end{adjustbox}
\end{table}

Across all approaches, the top-5 profitable SKUs were consistently selected in every planning period, reinforcing their importance as high-value anchors in the optimization process. Among the methods, Quanfluence and Quantum QUBO stood out for achieving the highest net profits, while PSO achieved more modest results but with lower costs and a narrower SKU portfolio.

\subsection{Quanfluence Quantum QUBO}
The Quanfluence Quantum QUBO delivered exceptional performance across all key metrics in this multi-period production planning scenario. Operating over eight planning periods with a capacity constraint of 28,392 units per period, the system processed a QUBO formulation containing over one million terms and approximately 4,100 decision variables. The optimizer achieved a best-found energy of $-2.95 \times 10^{16}$ and selected 288 distinct SKUs totaling 226,813 units, generating a net profit of \$12,752,661.09 against a scaled cost of \$187,182.28. This profit figure represents the highest value obtained across all competing methods in this study.
When compared against alternative approaches, Quanfluence Quantum consistently outperformed both quantum-inspired variants and classical metaheuristics. The results surpassed those from Quanfluence Cosine QUBO, SQA Quantum QUBO, and SQA Cosine QUBO configurations, while also exceeding the performance of established classical algorithms including Particle Swarm Optimization, Genetic Algorithm, and Ant Colony Optimization. Beyond raw profit maximization, the solution maintained substantial product diversity by selecting nearly 60\% of the available catalog, demonstrating that the coherent Ising machine approach effectively balances profitability with practical supply chain considerations such as market coverage and inventory distribution.

\begin{figure}[H]
    \centering
    \begin{minipage}{0.55\textwidth}
        \centering
        \includegraphics[width=\linewidth]{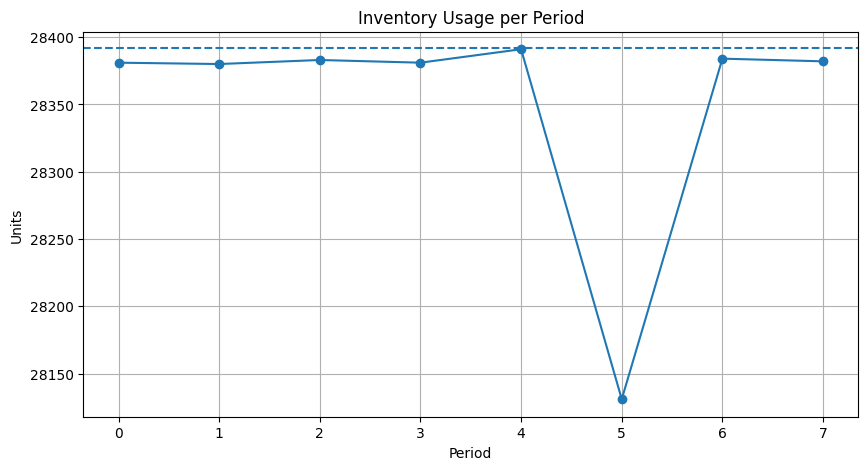}
    \end{minipage}\hfill
    \begin{minipage}{0.42\textwidth}
        \centering
        \includegraphics[width=\linewidth]{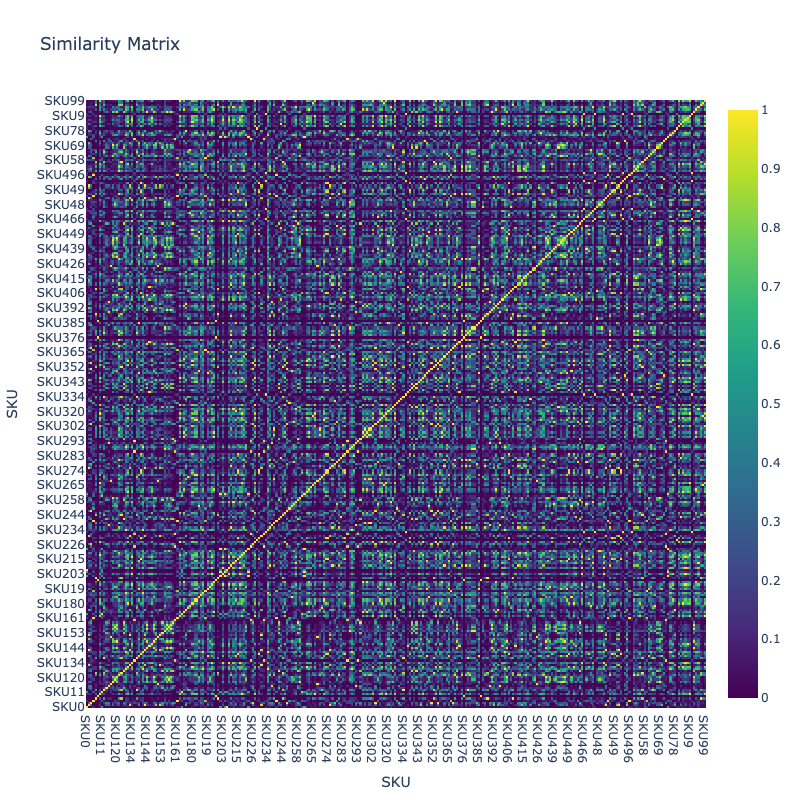}
    \end{minipage}
    \caption{Quanfluence quantum QUBO results: (left) multi-period capacity utilization; (right) quantum similarity matrix capturing SKU interdependencies.}
    \label{fig:quanfluence}
\end{figure}

\vspace{0.5em}
\noindent To provide a visual perspective, Figure~\ref{fig:quanfluence} presents the per-period capacity utilization and the quantum similarity matrix side by side. Together, these plots illustrate how the CIM distributed units across periods while preserving SKU interrelationships, offering both an operational and structural view of the optimization outcome.

\subsection{Quanfluence Cosine QUBO}
\begin{figure}[H]
    \centering
    \begin{minipage}{0.55\textwidth}
        \centering
        \includegraphics[width=\linewidth]{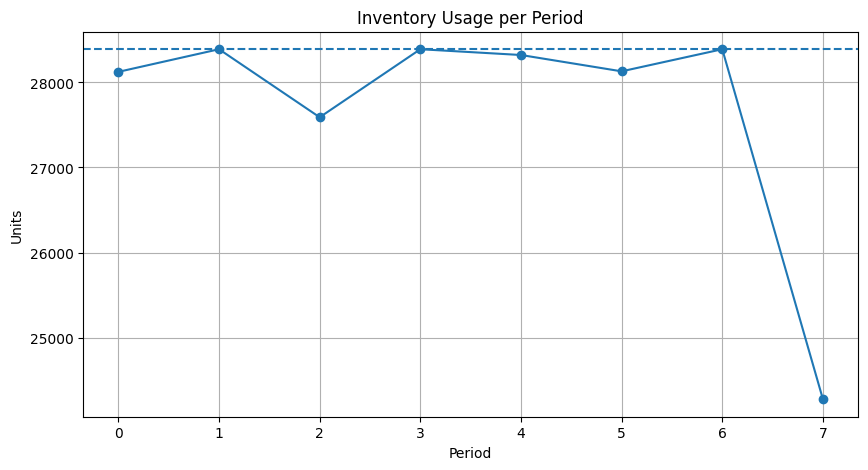}
    \end{minipage}\hfill
    \begin{minipage}{0.42\textwidth}
        \centering
        \includegraphics[width=\linewidth]{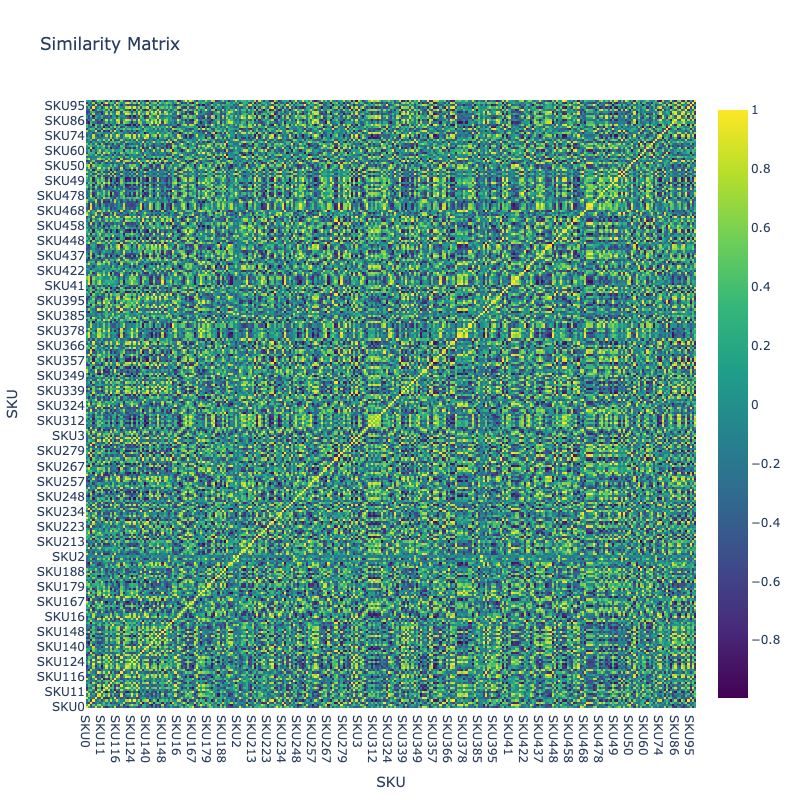}
    \end{minipage}
    \caption{Quanfluence cosine QUBO results showing capacity allocation across periods (left) and the cosine similarity matrix between SKUs (right).}
    \label{fig:cosine_quanfluence}
\end{figure}

The Quanfluence approach using cosine similarity and QUBO formulation delivered impressive results, selecting 283 distinct SKUs across all periods and covering 221,611 units. The method generated a net profit of \$12,210,262.79 with a total cost of \$172,672.61, working within a capacity constraint of 28,392 units per period. What's particularly noteworthy is that it maintained the top-5 most profitable SKUs—SKU308, SKU439, SKU47, SKU250, and SKU11—across every single period, showing remarkable consistency in identifying and retaining high-value products.
This performance represents a significant jump from the traditional metaheuristics. While GA achieved around \$11.3 million and ACO reached \$10.5 million, Quanfluence pushed past \$12.2 million by combining quantum-inspired optimization with the broader exploration characteristics of cosine similarity. The selection of 283 SKUs sits right between GA's 271 and ACO's 277, but the unit coverage and profit suggest Quanfluence made smarter choices about which products to include and how to allocate capacity across periods. The cosine similarity matrix (Figure \ref{fig:cosine_quanfluence}, right) shows the characteristic dense connectivity pattern we saw with GA and ACO, indicating that Quanfluence explored a wide range of SKU relationships rather than being overly selective.
The multi-period optimization capability proved valuable here, as shown in the capacity utilization pattern (Figure \ref{fig:cosine_quanfluence}, left). Rather than making a single static selection, Quanfluence distributed the 221,611 units strategically across periods while respecting capacity limits. This dynamic allocation likely captured seasonal patterns or demand fluctuations that single-period approaches would miss, contributing to the higher overall profitability while maintaining operational feasibility.

\subsection{Quantum QUBO Results}
\begin{figure}[H]
    \centering
    \begin{minipage}{0.55\textwidth}
        \centering
        \includegraphics[width=\linewidth]{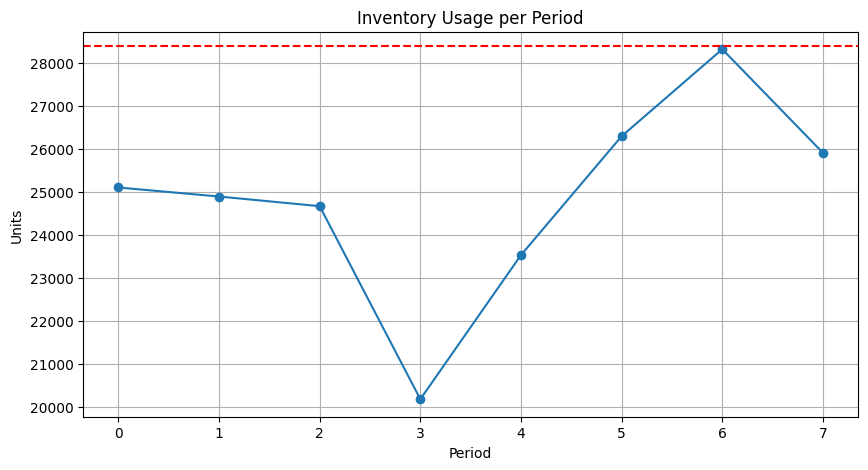}
    \end{minipage}\hfill
    \begin{minipage}{0.42\textwidth}
        \centering
        \includegraphics[width=\linewidth]{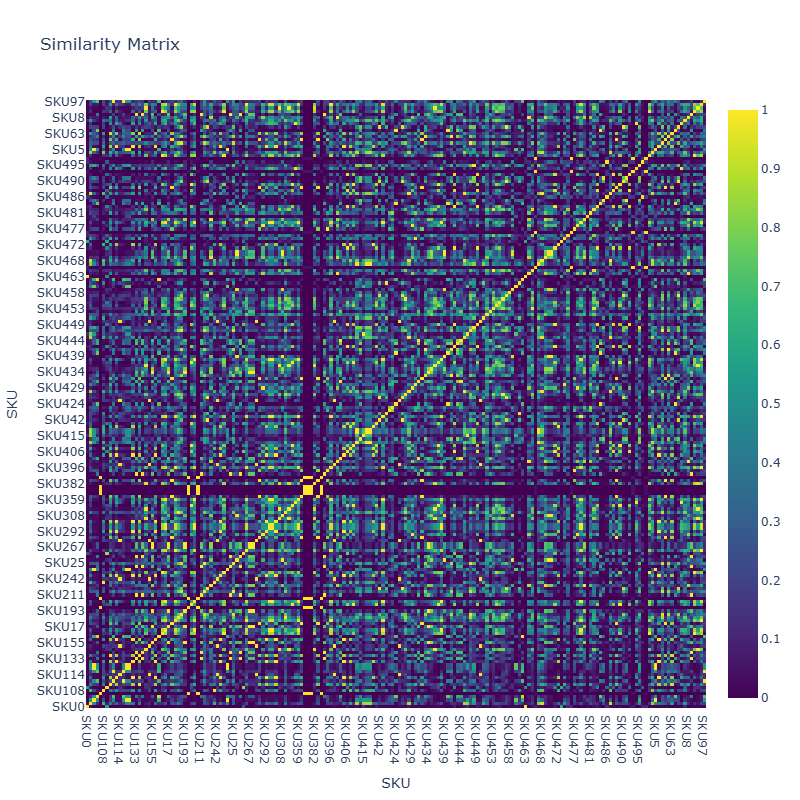}
    \end{minipage}
    \caption{SQA cosine QUBO results showing capacity allocation across periods (left) and the cosine similarity matrix between SKUs (right).}
    \label{fig:quantum_SQA}
\end{figure}

The Simulated Quantum Annealing approach using quantum similarity delivered a notably refined performance, selecting 191 SKUs and covering 198,907 units. It generated a net profit of \$11,894,758.25 with a total cost of \$157,290.07, maintaining the same 28,392 units per period capacity constraint. Just like its cosine counterpart, SQA with quantum similarity locked in all top-5 profitable SKUs—SKU308, SKU439, SKU47, SKU250, and SKU11—across every period, proving its reliability in identifying core high-value products.
The jump from SQA cosine to SQA quantum is subtle but meaningful. With just 11 additional SKUs (191 versus 180) and slightly higher unit coverage (198,907 versus 194,041), the quantum version squeezed out an extra \$194,000 in profit. This improvement comes from the discriminative power of quantum similarity, which you can see in the matrix (Figure \ref{fig:quantum_SQA}, right). Unlike the dense, broadly connected cosine patterns we've seen before, the quantum similarity matrix shows much sparser relationships with clearer boundaries between similar and dissimilar SKUs. This helped SQA make smarter additions to the portfolio—those 11 extra SKUs weren't just filling space, they were genuinely complementary products that added real value.
Looking at the capacity allocation (Figure \ref{fig:quantum_SQA}, left), SQA quantum managed its slightly larger portfolio efficiently across periods, distributing units in a way that respected constraints while maximizing returns. The annealing process combined with quantum similarity seems to have found a balanced middle ground—more selective than Quanfluence's 283 SKUs but slightly more expansive than SQA cosine's 180, landing on 191 as the optimal portfolio size. This sweet spot delivered profits that nearly matched Quanfluence (\$11.89M versus \$12.21M) while keeping costs lower and the portfolio more manageable.

\subsection{SQA Cosine QUBO Results}
\begin{figure}[H]
    \centering
    \begin{minipage}{0.55\textwidth}
        \centering
        \includegraphics[width=\linewidth]{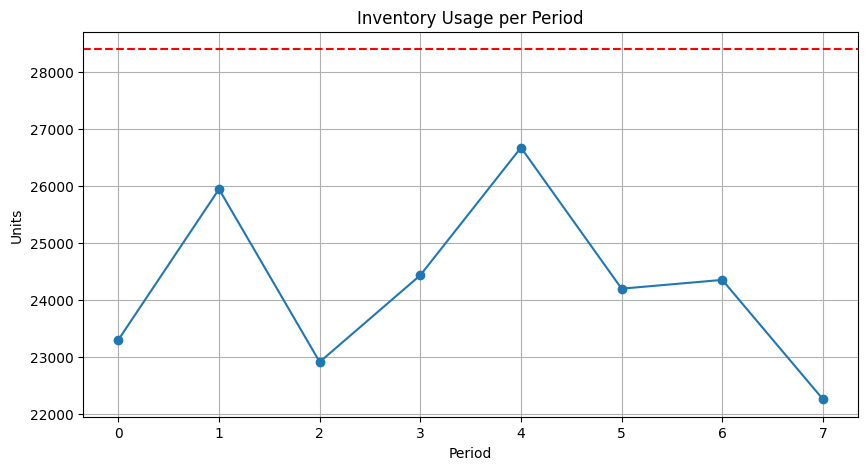}
    \end{minipage}\hfill
    \begin{minipage}{0.42\textwidth}
        \centering
        \includegraphics[width=\linewidth]{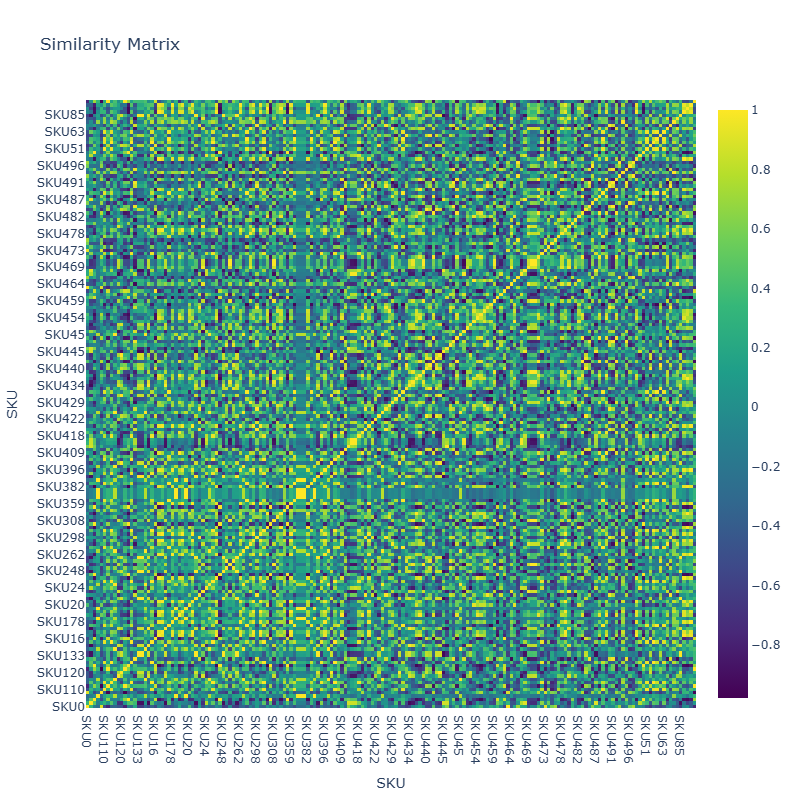}
    \end{minipage}
    \caption{SQA cosine QUBO results showing capacity allocation across periods (left) and the cosine similarity matrix between SKUs (right).}
    \label{fig:cosine_SQA}
\end{figure}

The Simulated Quantum Annealing approach with cosine similarity took a noticeably different path compared to Quanfluence, selecting just 180 SKUs and covering 194,041 units. Despite this more conservative selection, SQA still delivered strong results with a net profit of \$11,700,278.44 and a total cost of \$154,947.81, all while working within the same 28,392 units per period capacity constraint. Like Quanfluence, it consistently maintained all top-5 profitable SKUs—SKU308, SKU439, SKU47, SKU250, and SKU11—throughout every period, confirming its ability to lock onto genuinely valuable products.
What's interesting here is the tradeoff SQA made. With 180 SKUs compared to Quanfluence's 283, it was far more selective about what to include, yet it still achieved \$11.7 million in profit—only about \$500,000 less than Quanfluence. This suggests SQA found a sweet spot between portfolio size and profitability. Looking at the similarity matrix (Figure \ref{fig:cosine_SQA}, right), you can see the familiar dense pattern of cosine similarity, but SQA's optimization process clearly filtered through these connections more aggressively, keeping only the combinations that truly contributed to the bottom line rather than just adding SKUs for the sake of coverage.
The capacity utilization across periods (Figure \ref{fig:cosine_SQA}, left) shows how SQA balanced its smaller SKU portfolio over time. Even with fewer products to work with, it managed to distribute 194,041 units strategically, staying well within capacity limits while maintaining profitability. This efficiency is particularly impressive—SQA achieved nearly the same profit as GA (\$11.3 million) but did so with 91 fewer SKUs and lower overall costs. The annealing process seems to have helped it avoid the bloat that sometimes comes with evolutionary approaches, settling into a lean, high-performing portfolio instead.

\subsection{PSO Results}
The PSO optimization using cosine similarity achieved a net profit of \$4,079,022.82 from 83 SKUs, which was considerably lower than what the quantum approach delivered. Interestingly, PSO did manage to consistently pick the top-5 most profitable SKUs, so it wasn't making poor choices at the individual product level. The problem was more subtle than that.
Cosine similarity works by measuring the angle between product feature vectors, and this tends to make lots of products look reasonably similar to each other. Looking at the similarity matrix, you can see this in how the colors are spread across yellows, greens, and teals - there aren't many sharp distinctions. Everything seems moderately related to everything else, with gradual transitions between similar and dissimilar pairs.
This created a real challenge for the optimization. Sure, PSO found the high-margin winners, but it struggled to understand which products actually work well together as a portfolio. The algorithm ended up selecting 83 SKUs that individually made sense but probably had overlapping appeal to the same customers. 

\begin{figure}[H]
    \centering
    \includegraphics[width=0.55\textwidth]{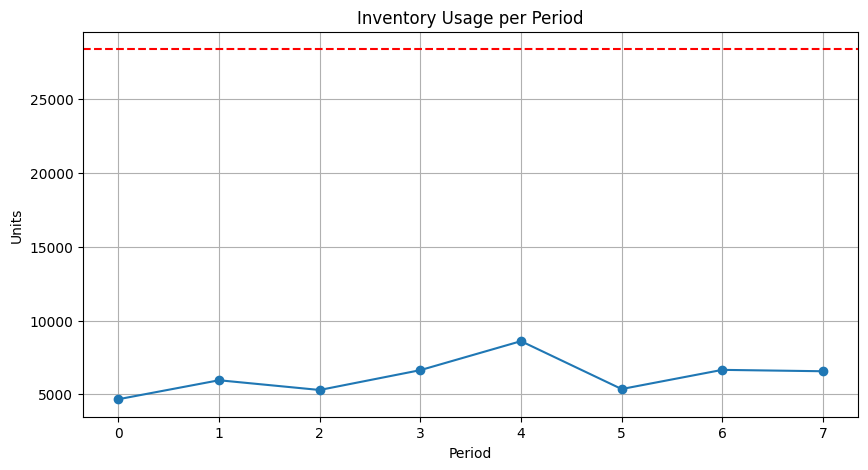}
    \hfill
    \includegraphics[width=0.42\textwidth]{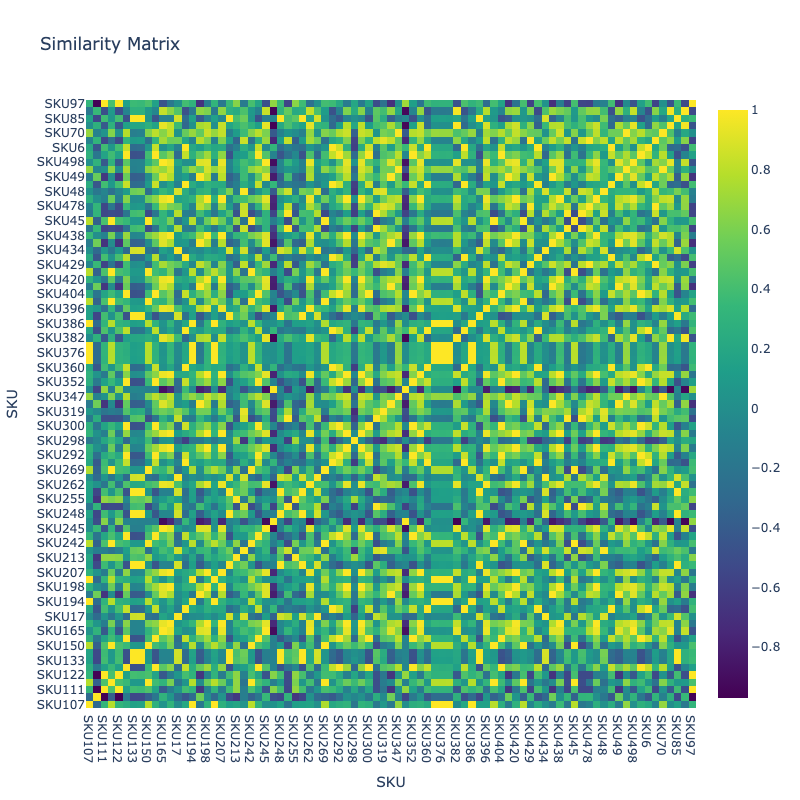}
    \caption{PSO: Capacity utilization across periods (left) and similarity matrix of SKU selection (right).}
    \label{fig:pso-results}
\end{figure}

Figures~\ref{fig:pso-results} show capacity usage (left) and SKU similarity (right). The underutilization of capacity is visible, yet stability in SKU overlap remains evident.  

\subsection{Genetic Algorithm and Ant Colony Optimization Results}

\begin{figure}[H]
    \centering
    \subfloat[GA: Capacity utilization across periods]{%
        \includegraphics[width=0.55\textwidth]{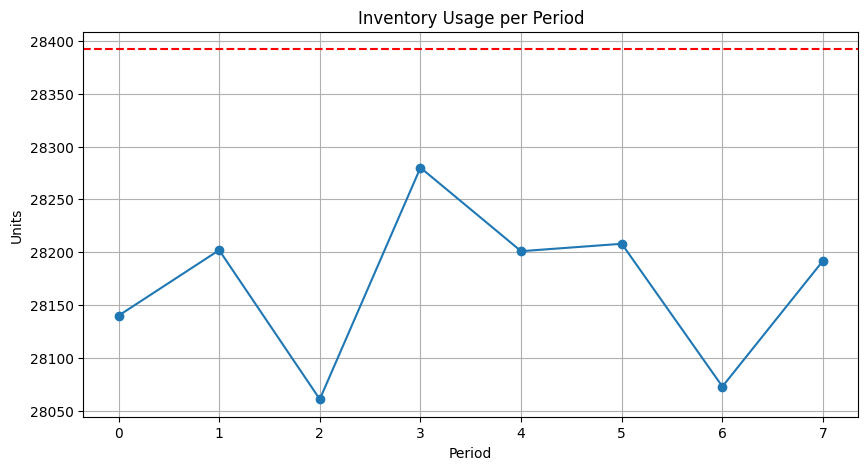}
        \label{fig:ga-capacity}
    }
    \hfill
    \subfloat[GA: Similarity matrix of SKU selection]{%
        \includegraphics[width=0.42\textwidth]{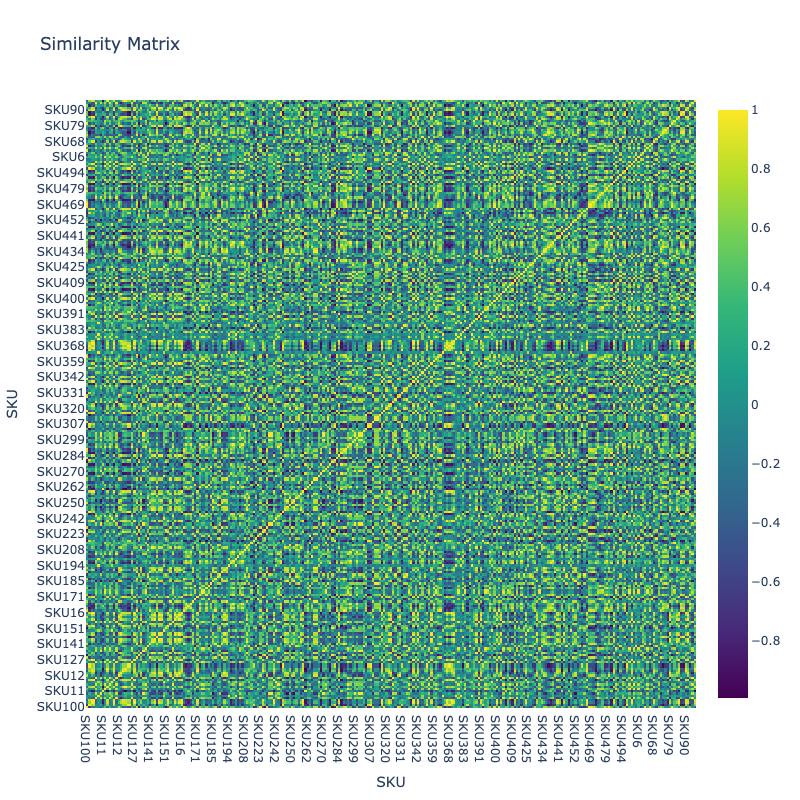}
        \label{fig:ga-similarity}
    }
    
    \vspace{0.5cm}
    
    \subfloat[ACO: Capacity utilization across periods]{%
        \includegraphics[width=0.55\textwidth]{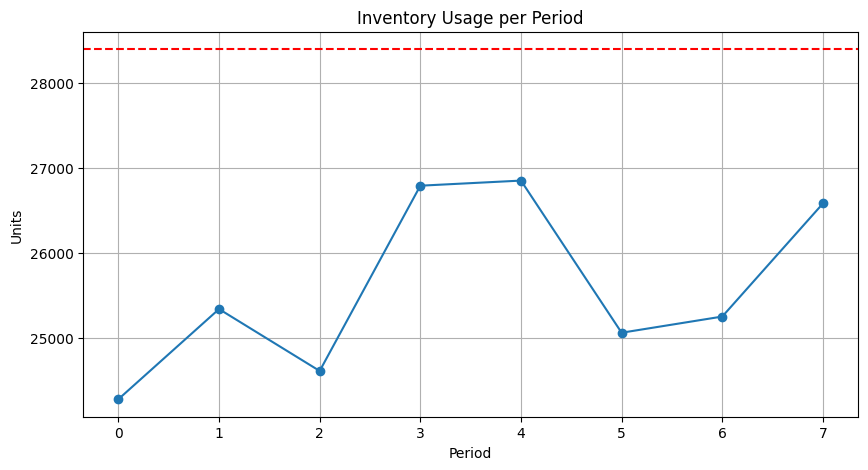}
        \label{fig:aco-capacity}
    }
    \hfill
    \subfloat[ACO: Similarity matrix of SKU selection]{%
        \includegraphics[width=0.42\textwidth]{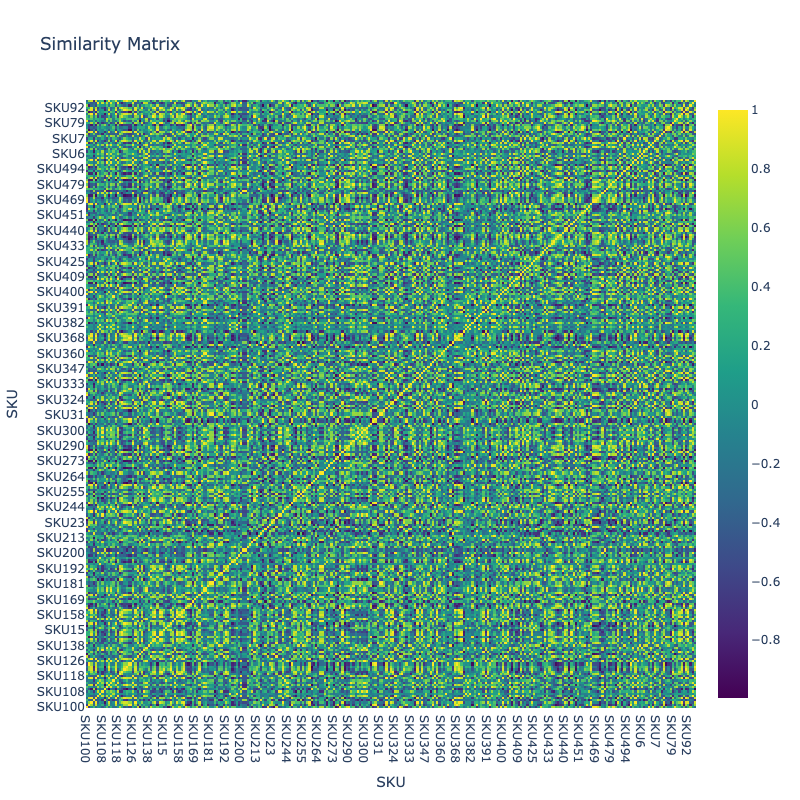}
        \label{fig:aco-similarity}
    }

    \caption{Comparison of GA and ACO performance: capacity utilization (left) and similarity matrices (right).}
    \label{fig:ga-aco-results}
\end{figure}

Both GA and ACO performed remarkably well compared to PSO, though they took notably different approaches to building their portfolios. The Genetic Algorithm selected 271 SKUs covering 225,357 units and generated a net profit of \$11,304,904.34 with a total cost of \$168,586.61 (Figure \ref{fig:ga-capacity}). ACO wasn't far behind, selecting 277 SKUs covering 204,766 units and achieving a net profit of \$10,518,980.76 at a cost of \$160,129.82 (Figure \ref{fig:aco-capacity}). What's particularly reassuring is that both methods consistently picked the top-5 most profitable SKUs, showing they weren't just randomly exploring the solution space but actually honing in on genuine high-performers.
The profit gap between GA and ACO—roughly \$786,000 in GA's favor—seems to come down to coverage strategy. GA managed to cover more units (225,357) with fewer SKUs (271), suggesting it found combinations that appealed to broader customer segments as evident in the capacity allocation patterns (Figures \ref{fig:ga-capacity} and \ref{fig:aco-capacity}). ACO selected six more SKUs but covered fewer units overall, which hints that it may have gotten drawn toward some niche products that looked promising individually but didn't expand market reach as effectively. The similarity matrices (Figures \ref{fig:ga-similarity} and \ref{fig:aco-similarity}) reveal the dense connectivity patterns underlying both approaches. This is probably a consequence of how pheromone trails work in ACO—once the algorithm finds a decent path, it tends to reinforce it, which can sometimes mean missing slightly better alternatives.
What really stands out is how much both methods outperformed PSO, which only managed about \$4 million in profit with 83 SKUs. GA and ACO achieved roughly 2.5x to nearly 3x that profit by casting a much wider net. The capacity utilization and similarity patterns (Figure \ref{fig:ga-aco-results}). They weren't being overly selective about which SKUs to include; instead, they explored larger portfolios that balanced diversity with profitability. This broader exploration paid off handsomely, capturing customer demand across multiple segments rather than focusing narrowly on just the obvious winners.

\section{Ablation Study}

An ablation study evaluates the contribution of individual QUBO components by systematically removing them and observing the effect on feasibility, profit, and diversity. Each experiment was repeated five times with different random seeds, and results are reported as mean $\pm$ standard deviation. The complete model (\texttt{Quantum\_QUBO\_Full}) serves as the baseline, achieving feasible allocations with mean profit $12.68 \pm 0.25$M, coverage of $338 \pm 6$ SKUs, no capacity violations, and low redundancy.

\subsection{Setup and Notation}

Decision variables are binary selections $x_{t,i} \in \{0,1\}$ for SKU $i$ in period $t$. Per-period slack bits $s_{t,b}$ enforce capacity. Demand of SKU $i$ is $D_i$, unit margin is $U_i$, and total margin is $U_i D_i$. The QUBO formulation is

\begin{align*}
E(\mathbf{x},\mathbf{s}) &= 
-\lambda_m \sum_i U_i D_i x_i \\
&\quad + \lambda_{inv} \sum_i \mathrm{InvRisk}_i x_i 
+ \lambda_{def} \sum_i \mathrm{DefRisk}_i x_i 
+ \lambda_{risk} \sum_i \mathrm{Risk}_i D_i x_i \\
&\quad + \lambda_s \sum_{i<j} S_{ij} x_i x_j \\
&\quad + \lambda_c \sum_t \Bigg( \sum_i D_i x_{t,i} 
     + \sum_b 2^b s_{t,b} - C \Bigg)^2 \\
&\quad + \lambda_k \Big( \sum_i x_i - K \Big)^2 
- \lambda_{top5} \sum_{i \in \mathrm{Top5}} x_i ,
\end{align*}

where $S_{ij}$ denotes quantum similarity (PCA + embedding), $C$ is capacity per period, $K$ is the SKU count target, and $\lambda_{\cdot}$ are penalty weights.

\subsection{Experiment: NoCapacity}
\textbf{Term removed:} quadratic slack capacity term. \\
\textbf{Observed effect:} Profit rose to $25.50 \pm 0.07$M, but coverage collapsed to $110 \pm 2$ SKUs, with an average of 8 capacity violations. Redundancy increased sharply ($0.72 \pm 0.01$). \\
\textbf{Interpretation:} $\lambda_c$ is structural; without it, allocations are infeasible as the solver selects unconstrained high-demand SKUs.

\subsection{Experiment: NoSimilarity}
\textbf{Term removed:} similarity penalty $\lambda_s \sum_{i<j} S_{ij} x_i x_j$. \\
\textbf{Observed effect:} Profit remained close to baseline ($12.63 \pm 0.40$M), and coverage was stable ($340 \pm 4$ SKUs). However, redundancy increased ($11{,}750 \pm 378$ vs.\ $11{,}535 \pm 365$ baseline) and the redundancy score worsened ($-0.030 \pm 0.037$ vs.\ $-0.027 \pm 0.017$). \\
\textbf{Interpretation:} Profitability is largely unaffected, but assortment diversity declines. $S_{ij}$ is essential for reducing redundancy and maintaining balance.

\subsection{Experiment: NoMarginWeight}
\textbf{Term removed:} profit reward $-\lambda_m \sum_i U_i D_i x_i$. \\
\textbf{Observed effect:} Profit dropped to $12.54 \pm 0.18$M, SKU coverage remained $338 \pm 3$, and redundancy was comparable. \\
\textbf{Interpretation:} $\lambda_m$ is the principal profit driver; without it, the solver reduces emphasis on high-margin SKUs.

\subsection{Experiment: NoLegacyInvRisk / NoRisk}
\textbf{Terms removed:} $\lambda_{inv}, \lambda_{def}, \lambda_{risk}$. \\
\textbf{Observed effect:} Profit was $12.51 \pm 0.21$M, SKU coverage slightly increased to $346 \pm 5$, with marginally higher redundancy. \\
\textbf{Interpretation:} These penalties act as stabilizers, biasing toward safer SKUs but not dominating profitability.

\subsection{Experiment: NoSkuLimit}
\textbf{Term removed:} SKU-count spread penalty $\lambda_k (\sum_i x_i - K)^2$. \\
\textbf{Observed effect:} Profit increased to $12.82 \pm 0.16$M, SKU coverage was $341 \pm 3$, with redundancy unchanged. \\
\textbf{Interpretation:} $\lambda_k$ balances breadth versus concentration; removing it encourages concentrated high-margin selections.

\subsection{Experiment: NoTop5}
\textbf{Term removed:} top-SKU incentive $-\lambda_{top5}\sum_{i\in\mathrm{Top5}} x_i$. \\
\textbf{Observed effect:} Profit fell to $11.00 \pm 0.37$M, while SKU breadth increased to $364$. Redundancy worsened ($13{,}137 \pm 444$ pairs). \\
\textbf{Interpretation:} $\lambda_{top5}$ secures high-margin SKUs; its removal reduces profit despite broader coverage.

\subsection{Experiment: NoPCA}
\textbf{Change:} $S_{ij}$ computed without PCA. \\
\textbf{Observed effect:} Profit was $12.81 \pm 0.06$M, SKU coverage $341 \pm 5$, but redundancy increased with a positive redundancy score ($0.120 \pm 0.009$). \\
\textbf{Interpretation:} PCA enhances structure in $S_{ij}$; without it, similarity is noisier, reducing diversity.

\subsection{Summary}

\begin{table}[H]
\centering
\caption{Ablation summary with corrected NoSimilarity results}
\label{tab:ablation_results}
\small 
\begin{adjustbox}{max width=\textwidth}
\begin{tabular}{
    l
    r@{\hspace{4pt}}r
    r@{\hspace{4pt}}r
    r@{\hspace{4pt}}r
    r@{\hspace{4pt}}r
    r@{\hspace{4pt}}r
    r@{\hspace{4pt}}r
    r@{\hspace{4pt}}r
}
\toprule
\textbf{Experiment} &
\multicolumn{2}{c}{\textbf{Total Profit}} &
\multicolumn{2}{c}{\textbf{Total Units}} &
\multicolumn{2}{c}{\textbf{Distinct SKUs}} &
\multicolumn{2}{c}{\textbf{Cap. Violations}} &
\multicolumn{2}{c}{\textbf{Cap. Excess}} &
\multicolumn{2}{c}{\textbf{Redundant Pairs}} &
\multicolumn{2}{c}{\textbf{Avg. Redundancy}} \\
\cmidrule(lr){2-3}
\cmidrule(lr){4-5}
\cmidrule(lr){6-7}
\cmidrule(lr){8-9}
\cmidrule(lr){10-11}
\cmidrule(lr){12-13}
\cmidrule(lr){14-15}
& \textbf{Mean} & \textbf{Std} 
& \textbf{Mean} & \textbf{Std}
& \textbf{Mean} & \textbf{Std}
& \textbf{Mean} & \textbf{Std}
& \textbf{Mean} & \textbf{Std}
& \textbf{Mean} & \textbf{Std}
& \textbf{Mean} & \textbf{Std} \\
\midrule
Quantum\_QUBO\_Full & 12,679,277.65 & 251,034.53 & 224,465.80 & 909.96 & 338.00 & 5.79 & 0.00 & 0.00 & 0.00 & 0.00 & 11,535.00 & 365.20 & -0.03 & 0.02 \\
Quantum\_QUBO\_NoCapacity & 25,501,055.68 & 70,224.32 & 314,387.80 & 1,002.14 & 110.20 & 2.17 & 8.00 & 0.00 & 87,251.80 & 2,011.62 & 1,785.20 & 38.40 & 0.72 & 0.01 \\
Quantum\_QUBO\_NoSimilarity & 12,626,373.74 & 398,070.06 & 224,865.20 & 1,038.01 & 340.40 & 3.85 & 0.00 & 0.00 & 0.00 & 0.00 & 11,750.40 & 378.01 & -0.03 & 0.04 \\
Quantum\_QUBO\_NoMarginWeight & 12,540,445.95 & 177,095.73 & 224,873.40 & 989.59 & 338.00 & 2.55 & 0.00 & 0.00 & 0.00 & 0.00 & 11,443.80 & 423.28 & -0.03 & 0.02 \\
Quantum\_QUBO\_NoLegacyInvRisk & 12,509,353.69 & 210,773.57 & 225,300.00 & 1,107.10 & 346.00 & 4.53 & 0.00 & 0.00 & 0.00 & 0.00 & 12,045.40 & 352.26 & -0.03 & 0.01 \\
Quantum\_QUBO\_NoSkuLimit & 12,821,061.41 & 164,984.18 & 225,170.80 & 1,035.85 & 340.80 & 2.85 & 0.00 & 0.00 & 0.00 & 0.00 & 11,468.80 & 248.12 & -0.02 & 0.02 \\
Quantum\_QUBO\_NoTop5 & 10,997,235.52 & 367,142.65 & 223,637.00 & 1,223.16 & 364.00 & 0.00 & 0.00 & 0.00 & 0.00 & 0.00 & 13,137.20 & 443.85 & -0.02 & 0.01 \\
Quantum\_QUBO\_NoPCA & 12,813,485.89 & 60,815.46 & 224,632.00 & 1,120.25 & 341.20 & 4.80 & 0.00 & 0.00 & 0.00 & 0.00 & 11,444.20 & 403.11 & 0.12 & 0.01 \\
\bottomrule
\end{tabular}
\end{adjustbox}
\end{table}

\begin{itemize}
    \item \textbf{Structural terms:} $\lambda_c, \lambda_s, \lambda_{top5}$ are essential for feasibility and diversity. 
    \item \textbf{Profit drivers:} $\lambda_m$ and $\lambda_{top5}$ dominate profitability.
    \item \textbf{Stabilizers:} $\lambda_{inv}, \lambda_{def}, \lambda_{risk}, \lambda_k$ enhance robustness but have secondary effect.
    \item \textbf{Preprocessing:} PCA is critical to ensure $S_{ij}$ reflects meaningful correlations.
\end{itemize}

Overall, capacity, similarity, and top-SKU constraints are indispensable for feasible, diverse, and profitable solutions. Profit is primarily driven by margin and top-SKU incentives, while stabilizers reduce risk. PCA preprocessing improves similarity quality. These results, in Table~\ref{tab:ablation_results}, confirm that the complete QUBO formulation offers the most reliable trade-off between feasibility, profitability, and robustness.

\section{Conclusion}

This work has demonstrated that large-scale multi-period supply chain allocation can be expressed and solved within a unified QUBO framework that integrates profitability, risk control, and assortment diversity under strict operational capacity limits. By combining a quantum-derived similarity kernel with slack-bit capacity enforcement, the formulation was executed on a Coherent Ising Machine and benchmarked against classical and simulated quantum annealing baselines. The results show that physics-inspired optimizers can generate feasible allocations that simultaneously maximize profit and maintain SKU diversity, offering outcomes that are both interpretable and operationally meaningful.

At the same time, several practical limitations must be acknowledged. The similarity kernel was generated through simulated quantum embeddings rather than direct execution on near-term noisy hardware; capacity feasibility depends on careful penalty scaling; and variable counts still grow linearly with both SKU set size and planning horizon, constraining applicability to very large instances. These factors suggest that hybrid decomposition strategies, hierarchical SKU screening, or tighter integration with domain-specific heuristics will be necessary to extend scalability in industrial practice.

Looking forward, the framework presented here offers a blueprint for embedding domain-aware penalties, diversity controls, and exact feasibility constraints into QUBO formulations that can be solved on emerging quantum or quantum-inspired devices. As hardware platforms evolve, such formulations have the potential to move b1eyond proof-of-concept studies and directly inform resilient, profitable, and diversified supply chain planning in real-world operations.


\begin{thebibliography}{99}

\bibitem{ref1} M. Sharma, H. C. Lau. A Comparative Study of Quantum Optimization Techniques for Solving Combinatorial Optimization Benchmark Problems, 2025. Available at: \url{https://arxiv.org/abs/2503.12121v1}  
\bibitem{ref2} A. Morais, E. Osaba, I. Pastor, I. Oregui. Comparative Analysis of Classical and Quantum-Inspired Solvers: A Preliminary Study on the Weighted Max-Cut Problem, 2025. Available at: \url{https://arxiv.org/abs/2504.05989v1}  
\bibitem{ref3} A. Bochkarev, R. Heese, S. J\"{a}ger, P. Schiewe, A. Sch\"{o}bel. Quantum Computing for Discrete Optimization: A Highlight of Three Technologies, 2025. Available at: \url{https://arxiv.org/abs/2409.01373} 
\bibitem{refk} E. Crosson, A. W. Harrow. Simulated Quantum Annealing Can Be Exponentially Faster than Classical Simulated Annealing, 2016. Available at: \url{https://arxiv.org/abs/1601.03030} 
\bibitem{ref4} F. Phillipson. Quantum Computing in Logistics and Supply Chain Management - an Overview, 2025. Available at: \url{https://arxiv.org/abs/2402.17520v2}  
\bibitem{ref5} S. Katoch, S. S. Chauhan, V. Kumar. ``A review on genetic algorithm: past, present, and future.'' \textit{Multimed Tools Appl} 80, 8091–8126, 2021. \url{https://doi.org/10.1007/s11042-020-10139-6 } 
\bibitem{ref6} Y. Li, Z. Ruan. ``Optimization of Logistics Supply Chain Based on Genetic Algorithm and Evolutionary Game.'' \textit{Journal of Advanced Manufacturing Systems}. \url{https://doi.org/10.1142/S0219686726500253}
\bibitem{ref7} T. Li. ``Improvement to application of PSO algorithm on TSP problems.''  Applied and Computational Engineering 74 97-103, 2024. \url{https://doi.org/10.54254/2755-2721/74/20240446}  
\bibitem{ref8} R. S. Kadadevaramath, J. C. H. Chen, B. L. Shankar, K. Rameshkumar.
``Application of particle swarm intelligence algorithms in supply chain network architecture optimization.''
\textit{Expert Systems with Applications} 39(11), 10160--10176, 2012. \url{https://doi.org/10.1016/j.eswa.2012.02.116}  
\bibitem{ref9} M. Dorigo, M. Birattari, T. Stutzle. ``Ant colony optimization,'' \textit{IEEE Computational Intelligence Magazine}, 1(4), 28--39, 2006. \url{https://doi.org/10.1109/MCI.2006.329691}
\bibitem{ref10} Y. Liu.
``Application of Ant Colony Algorithm in Enterprise Supply Chain Network Optimization and Synergy Effect Analysis.''
\textit{Procedia Computer Science} 262, 201--207, 2025.
\url{https://doi.org/10.1016/j.procs.2025.05.045}  
\bibitem{ref11} J. B. Holliday, D. Blount, E. Osaba, K. Luu. Advanced Quantum Annealing Approach to Vehicle Routing Problems with Time Windows, 2025 Available at: \url{https://arxiv.org/abs/2503.24285v1}  
\bibitem{ref12} S. Sinno, T. Gro\ss{}, A. Mott, A. Sahoo, D. Honnalli, S. Thuravakkath, B. Bhalgamiya. Performance of Commercial Quantum Annealing Solvers for the Capacitated Vehicle Routing Problem, 2023. Available at: \url{https://arxiv.org/abs/2309.05564}  
\bibitem{ref13} M. Q. Mohammed, H. Mee\ss{}, M. Otte. ``Review of the application of quantum annealing-related technologies in transportation optimization.'' \textit{Quantum Inf Process} 24, 296, 2025. \url{https://doi.org/10.1007/s11128-025-04870-y}
\bibitem{ref14} R. Hamerly, \textit{et al.} ``Experimental investigation of performance differences between coherent Ising machines and a quantum annealer.'' \textit{Sci. Adv.} 5,eaau0823, 2019. \url{https://doi.org/10.1126/sciadv.aau0823}  

\bibitem{amirmotefaker2021}
A. Motefaker, ``Supply Chain Dataset,'' Kaggle, 2021. [Online]. Available: \url{https://www.kaggle.com/datasets/amirmotefaker/supply-chain-dataset}

\bibitem{lucas2014}
A. Lucas, 
``Ising formulations of many NP problems,'' 
\textit{Frontiers in Physics}, 2014.

\bibitem{havlicek2019}
V. Havlíček, A. D. Córcoles, K. Temme, A. W. Harrow, A. Kandala, J. M. Chow, and J. M. Gambetta, 
``Supervised learning with quantum-enhanced feature spaces,'' 
\textit{Nature}, 2019.

\bibitem{inagaki2016}
T. Inagaki \textit{et al.}, 
``A coherent Ising machine for 2000-node optimization problems,'' 
\textit{Science}, 2016.
\bibitem{mcmahon2016}
P. L. McMahon, A. Marandi, Y. Haribara, R. Hamerly, C. Langrock, S. Tamate, T. Inagaki, H. Takesue, S. Utsunomiya, K. Aihara, R. L. Byer, M. M. Fejer, H. Mabuchi, Y. Yamamoto,  
“A fully programmable 100-spin coherent Ising machine with all-to-all connections,‘‘  
\textit{Science}, vol. 354, no. 6312, pp. 614–617, 2016.  

\bibitem{neukart2017}
F. Neukart, G. Compostella, C. Seidel, D. von Dollen, S. Yarkoni, and B. Parney, 
``Traffic flow optimization using a quantum annealer,'' 
\textit{Scientific Reports / arXiv}, 2017.

\bibitem{dwave2025}
D-Wave Systems Inc., 
``New Hybrid Solver: Constrained Quadratic Model,'' 
D-Wave Support / Whitepaper (CQM Hybrid), 2025.

\bibitem{whitepaper2021}
D-Wave Systems Inc., 
``A hybrid solver for constrained quadratic models (whitepaper),'' 
2021.
.  



\bibitem{pennylane} Bergholm, V., et al. (2018). PennyLane: Automatic differentiation of hybrid quantum-classical computations. arXiv preprint arXiv:1811.04968. Available at: \url{https://arxiv.org/abs/1811.04968}

\bibitem{lightning} Asadi, F., et al. (2022). Lightning-fast quantum circuit simulation with GPU acceleration. arXiv preprint arXiv:2206.11273. Available at: \url{https://arxiv.org/abs/2206.11273}

\bibitem{havlicek} Havlíček, V., et al. (2019). Supervised learning with quantum-enhanced feature spaces. Nature, 567(7747), 209-212. Available at: \url{https://doi.org/10.1038/s41586-019-0980-2}

\bibitem{schuld} Schuld, M., \& Killoran, N. (2019). Quantum machine learning in feature Hilbert spaces. Physical Review Letters, 122(4), 040504. Available at: \url{https://doi.org/10.1103/PhysRevLett.122.040504}

\bibitem{normalization_review} Singh, D., \& Singh, B. (2020). Investigating the impact of data normalization on classification performance. Applied Soft Computing, 97, 105524. Available at: \url{https://doi.org/10.1016/j.asoc.2019.105524}

\bibitem{pca_ml} Lever, J., Krzywinski, M., \& Altman, N. (2017). Principal component analysis. Nature Methods, 14(7), 641-642. Available at: \url{https://doi.org/10.1038/nmeth.4346}

\bibitem{openjij} Zaman, M., Tanahashi, K., \& Tanaka, S. (2021). PyQUBO: Python Library for Mapping Combinatorial Optimization Problems to QUBO Form. IEEE Transactions on Computers, 71(4), 838-850. Available at: \url{https://github.com/OpenJij/OpenJij}

\bibitem{openjij_framework} Okada, S., et al. (2019). OpenJij: Framework for the Ising model and QUBO. Available at: \url{https://openjij.github.io/OpenJij/}

\bibitem{quanfluence} Quanfluence. Time-Multiplexed Coherent Ising Machine. Available at: \url{https://quanfluence.com/time-multiplexed-coherent-ising-machine/}















\end{thebibliography}
\end{document}